\documentclass[%
 reprint,
superscriptaddress,
 amsmath,amssymb,
 aps,
 pre,
]{revtex4-1}

\usepackage{graphicx}
\usepackage{dcolumn}
\usepackage{bm}
\usepackage[colorlinks= true, linkcolor=blue, citecolor=blue, urlcolor=blue]{hyperref}
\usepackage{lipsum}
\usepackage{bbold}
\usepackage{mathtools}
\usepackage[normalem]{ulem}
\usepackage{float}
\usepackage{adjustbox}
\usepackage{cancel}

\def\bea{\begin{eqnarray}}
\def\eea{\end{eqnarray}}

\usepackage{xcolor}

\newcommand{\eref}[1]{Eq.~(\ref{#1})}
\newcommand{\fref}[1]{Fig.~\ref{#1}} 

\begin{document}

\title{Target search optimization by threshold resetting}

\author{Arup Biswas}
\email{arupb@imsc.res.in}
\affiliation{The Institute of Mathematical Sciences, CIT Campus, Taramani, Chennai 600113, India}
\affiliation{Homi Bhabha National Institute, Training School Complex, Anushakti Nagar, Mumbai 400094, India}
\author{Satya N Majumdar}
\email{satyanarayan.majumdar@cnrs.fr}
\affiliation{Laboratoire de Physique Théorique et Modèles Statistiques (LPTMS), CNRS, Univ. Paris-Sud, Universite Paris-Saclay, 91405 Orsay, France}

\author{Arnab Pal}
\email{arnabpal@imsc.res.in}
\affiliation{The Institute of Mathematical Sciences, CIT Campus, Taramani, Chennai 600113, India}
\affiliation{Homi Bhabha National Institute, Training School Complex, Anushakti Nagar, Mumbai 400094, India}


\begin{abstract} 
We introduce a new class of first passage time optimization driven by threshold resetting, inspired by many natural processes where crossing a critical limit triggers failure, degradation or transition. In here, search agents are collectively reset when a threshold is reached, creating event-driven, system-coupled simultaneous resets that induce long-range interactions. We develop a unified framework to compute search times for these correlated stochastic processes, with ballistic- and diffusive searchers as key examples uncovering diverse optimization behaviors. A cost function, akin to breakdown penalties, reveals that optimal resetting can forestall larger losses. This formalism generalizes to broader stochastic systems with multiple degrees of freedom.
\end{abstract}

\pacs{Valid PACS appear here}
\begin{titlepage}
\maketitle
\end{titlepage}

\emph{\textbf{Introduction.}---}  Resetting mediated target search has been a focal point of scientific interest in statistical physics and stochastic processes. A hallmark feature of  resetting is that it can expedite the completion of complex processes by periodically reinstating a search process to a known or random state, thus uncovering new pathways, avoiding potential obstacles and in effect, enhancing the efficiency in locating desired targets \cite{evans_stochastic_2020,pal2024random}. The advent of resetting based optimization techniques has truly invigorated much theoretical \cite{evans_diffusion_2011,kusmierz2014first,pal2016diffusion,reuveni_optimal_2016,pal_first_2017,chechkin2018random,belan2018restart,pal2019first,kumar2023universal,de2023resetting,pal2019landau,sar2023resetting,bonomo2021mitigating,de2022optimal,bhat2016stochastic,campos2015phase,pal_search_2020,biswas2024search,nagar2016diffusion}, experimental \cite{tal2020experimental,besga2020optimal,faisant2021optimal,altshuler2024environmental,paramanick2024uncovering} and computational \cite{jolakoski2023first,blumer2024combining,munoz2025learning} research of first-passage phenomena.

The aim of this work is to introduce \textit{target search optimization} under a \textit{threshold resetting} (TR) mechanism. In contrast to the prototypical scenario where resetting is imposed by an external time clock, TR is an event driven  phenomenon in which resetting occurs when the system operates under “safety covenant” or broadly, meets certain ``threshold rules''. Thresholds play a crucial role in governing transitions across numerous systems, acting as critical points that trigger significant changes. In neuroscience, integrate-and-fire neurons accumulate stimuli until their membrane potential crosses a threshold, leading to an action potential before resetting to their initial state \cite{burkitt2006review,bachar2012stochastic,gerstein1964random,taillefumier2013phase}. Financial strategies like stop-loss and take-profit rely on preset thresholds to limit losses or secure gains \cite{kaufman2013trading,shiryaev2007optimal,zhang2001stock,miller1966model}. In software engineering, circuit breakers operate similar to threshold-based reset mechanisms, preventing an application to access an unresponsive server \cite{nygard2018release,surendro2021circuit,montesi2018decorator}. In physics, fiber bundle models describe how materials distribute stress until a rupture threshold is met, after which the load is redistributed among the remaining fibers \cite{pradhan2010failure,hansen2015fiber}. Simple stroboscopic threshold mechanisms have also been  employed to obtain sustained temporal regularity in chaotic systems with applications to laser modeling \cite{sinha2001using,sinha1998dynamics,sinha1999computing,bhowmick2014targeting}.

In the recent past, De Bruyne\textit{ et al} investigated spatial properties of a single diffusing particle under TR mechanism (referred as \textit{first-passage resetting} there) disclosing dynamic nonstationarity of the process \cite{de2020optimization,de2021optimization}. Yet, another unexplored aspect of the problem is the \textit{target search optimization} -- the study of which is at the heart of this letter. To put things in perspective, 
consider a first passage process that is conducted by a group of agents, setting off from a certain location
and looking for a target until it is found \cite{redner2001,bray2013persistence,metzler_first-passage_2014,martinez2013optimizing,bhattacharya2014collective,falcon2019collective}. The process is,
however, restarted at some random time whenever one of the agents reaches a pre-assigned threshold. At that time all the searchers are reset simultaneously to the initial locations from which they renew their search. This collective resetting couples the degrees of freedom and makes the system
strongly correlated  \cite{biroli2023extreme,biroli2023critical,biroli2024exact}. Such collective TR
mechanism has natural justifications as earlier instances of experiments with fish school \cite{couzin2005effective,couzin2011uninformed,conradt2005consensus} and swarming robots \cite{brambilla2013swarm,lerman2004review} have demonstrated event-triggered collective transitions—where a local individual event (such as the detection of a fault or hazard by a single agent) prompts a coordinated adjustment in navigation strategy or “reset” across
the population. Similarly, in individuals or groups of agents such as clonal raider ants (\textit{Ooceraea biroi}), sensory thresholds shaped by factors like environment, and external stimuli influence decisions vital for survival and reproduction  \cite{chase2025physics}. Inspired by these connections, this letter aims to analyze the statistical signatures of the random completion time for this correlated multi-agent search process under threshold controlled resetting. To this end, we develop a unified framework for generic governing dynamics and show that TR can also be used as a potent mechanism to facilitate improved target search dynamics and efficiency. Finally, inspired from the control-management problem, we  define a cost objective function that interpolates between the successful search time and the effective penalty owing to resetting events and examine whether this can be minimized. 

\begin{figure}
    \centering
    \includegraphics[width=8.5cm]{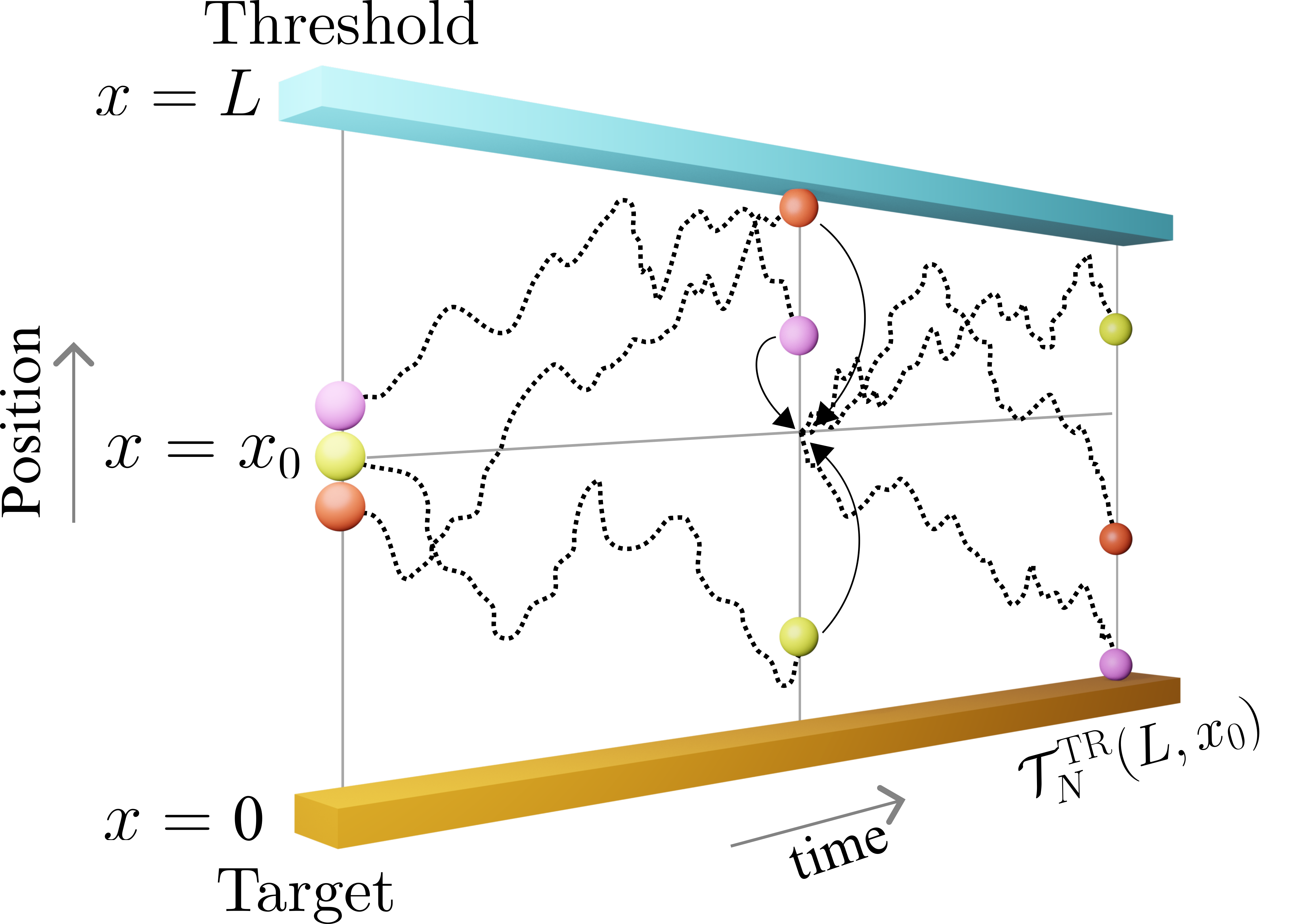}
    \caption{
    Schematic representation of the TR mechanism with $N$ searchers (here, $N=3$) starting from $x_0$ and undergoing generic stochastic dynamics. The process is considered complete when any of the searchers reaches the target located at the origin, and the corresponding first passage time is denoted by $ \mathcal{T}_N^\text{TR}(L,x_0)$. However, if any searcher reaches the resetting threshold at position 
$L$ before the target is found, all searchers are collectively returned to their initial positions, and the search process is restarted.}
    \label{fig1}
\end{figure}

\emph{\textbf{General formalism.}---} Consider a system of $N$ non-interacting searchers, each governed 
by the same stochastic dynamics which is generic, with position coordinates $\{ x_i \}$, starting from $x_0$. The searchers are confined in an interval $[0,L]$ where origin is  the target location.  If any of the $N$ searchers reaches this target we mark the process complete. The TR mechanism is introduced by placing another boundary at $L$, which hereafter is referred to as the \textit{threshold}  (see \fref{fig1}). Whenever one of the searchers hits the threshold, all of them are collectively reset to the initial position $x_0$ from where the search is renewed -- for a higher dimensional embedding interpretation, see \footnote{One natural way to interpret collective resetting is through the dynamics of a single ballistic searcher in $N$ dimensions, initialized at the point $(x_0, x_0, \ldots, x_0)$. For simplicity, let us consider the case $N = 2$. Imagine a square domain $(0, L) \times (0, L)$ with $0 \le x_0 \le L$. The target is located on two adjacent sides of the square: (1) $x_2 = 0$ with $0 \le x_1 \le L$, and (2) $x_1 = 0$ with $0 \le x_2 \le L$. The threshold, which triggers resetting, is defined by the other two sides: (3) $x_1 = L$ with $0 \le x_2 \le L$, and (4) $x_2 = L$ with $0 \le x_1 \le L$. A single particle begins at $(x_0, x_0)$ and moves with independently chosen velocities along the $x_1$ and $x_2$ directions. If it hits either of the threshold sides, it is reset to the initial point $(x_0, x_0)$ with new velocities — effectively resetting both coordinates \textit{simultaneously}. If it hits one of the target sides, the process ends. This setup allows us to interpret the dynamics of two independent 1D walkers (in our model) as the two components of a single 2D walker undergoing simultaneous resetting. This interpretation generalizes naturally to higher dimensions: $N$ independent walkers correspond to the $N$ components of a single walker in $N$ dimensions, with the simultaneous resetting occurring when any coordinate reaches its respective threshold. This provides an intuitive and unified perspective on the simultaneous resetting mechanism explored in our work. See also Fig. S11.
}. This procedure repeats until the target at the origin is found, with the process considered complete as soon as any one searcher locates it. We show that despite the presence of strong correlations, the mean search/first-passage time (MFPT) can be exactly calculated and furthermore, a nontrivial optimization is discovered due to the threshold resetting and the emerging collective nature of the searchers. Note that the model assumes instantaneous information transfer among searchers: whenever one reaches the threshold, all reset simultaneously; likewise, when one finds the target, all are immediately informed and the search ends.

\begin{figure*}
    \centering
    \includegraphics[width=17.7cm]{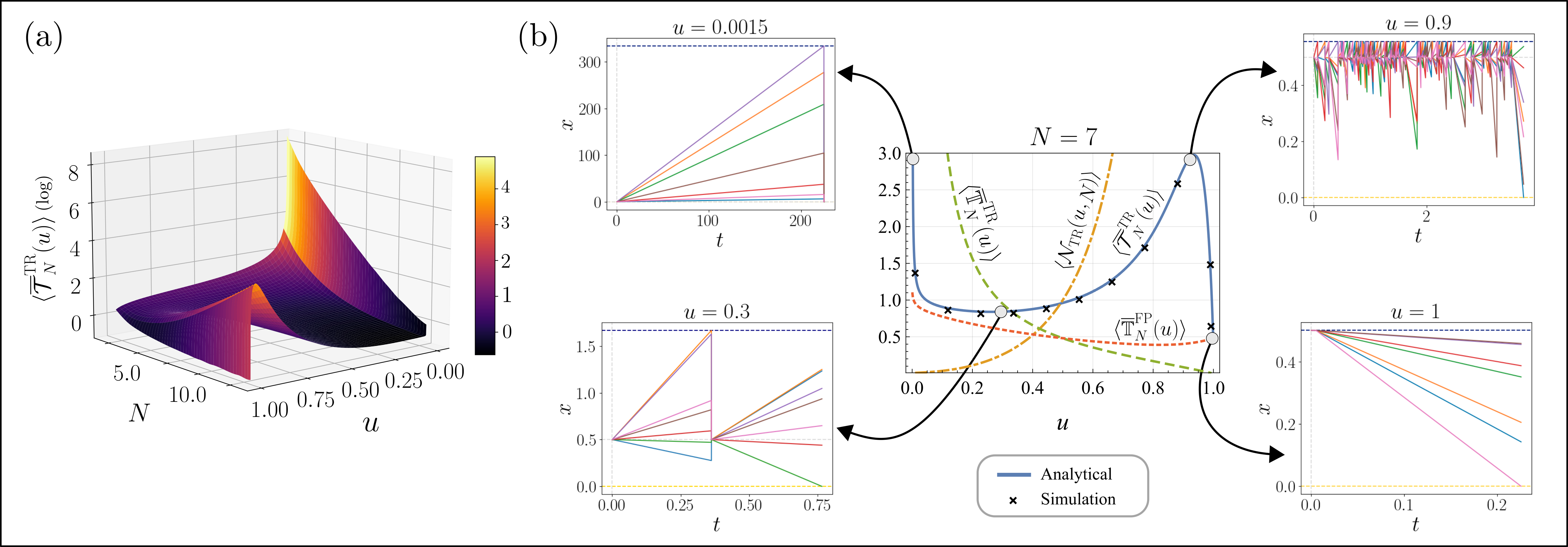}
    \caption{Panel (a): Variation of the scaled MFPT (\ref{scaling}) with respect to $u$ and $N$. The plot shows a global minima in the MFPT at $u=1$ for any $N>1$. A cross section for $N=7$ is shown in panel (b). Panel (b) - central figure: Optimization of $ \langle   \overline{\mathcal{T}}_N^\text{TR}(u) \rangle$ as a function of $u$ (shown by the solid blue curve) -- superimposed with individual components as given by the RHS of (\ref{fp-2}) namely $\langle {\mathcal{N}}_{\text{TR}} (u,N) \rangle$ (shown by orange dotted-dashed line), $ \langle \overline{\mathbb{T}}_N^{\text{TR}}(u) \rangle$ (shown by green dashed line) and $ \langle \overline{\mathbb{T}}_N^{\text{FP}}(u) \rangle$ (shown by red dotted line). Simulation results are shown by the cross markers demonstrating an excellent match. The accompanying subpanels showcase typical stochastic trajectories (with $x$ and $t$ standing for the position and time, respectively)} generated for different values of $u$ (marked by the circles) showing disparate behavior (global minima, local maxima \& minima, divergence) as explained in the main text. 
    In this analysis, we fix
    $x_0=0.5$ and vary $L$ within the range $[0.5,333.33]$ to explore the regime corresponding to $u \in (0.0015,1)$.  
    \label{traj}
\end{figure*}

We start by introducing the survival probability under TR as $\mathcal{Q}_N^\text{TR}(L,x_0,t)$ which estimates the probability that none of the $N$-particles has found the target at the origin, starting from $x_0$, up to time $t$ although they may remain confined without hitting the resetting threshold or they can hit the threshold once or multiple times. Summing over all these possibilities, we can write the following renewal equation for the survival probability 
\begin{eqnarray}
   &&\mathcal{Q}_N^\text{TR}(L,x_0,t)= \nonumber\\
   &&q_N(L,x_0,t)+\int_{0}^t dt' j_{L,N}(L,x_0,t') \mathcal{Q}_N^\text{TR}(L,x_0,t-t'),~~~~
   \label{surv-ren}
\end{eqnarray}
where $q_N(L,x_0,t)$ as the survival probability that none of the searchers has hit either of the boundaries at $x=0$ or $x=L$ up to time $t$
and $j_{L,N}(L,x_0,t')$ is the probability flux at the resetting threshold $L$ from any incoming searcher reaching at time $t'<t$. \eref{surv-ren} renders the following interpretation: Starting from $x_0$, the searchers can survive the target  upto time $t$ in two ways. In one plausible set of events, the searchers survive without hitting any of the boundaries which occurs with the probability $q_N(L,x_0,t)$ -- this is the ``no resetting'' scenario and the first term on the RHS. In the second case, whenever any one of the $N$ searchers hits the resetting threshold for the very \textit{first time} (say at $t'$), all are instantaneously reset to the starting position $x_0$ renewing the next trial for the search. The contribution for any particle hitting the resetting threshold at $L$ is essentially the flux $j_{L,N}(L,x_0,t')$ followed by the survival probability $\mathcal{Q}_N^\text{TR}(L,x_0,t-t')$ for the remaining duration. Taking the Laplace transform of \eref{surv-ren} with respect to $t$ and rearranging, we find
\begin{align}
    \widetilde{\mathcal{Q}}_N^\text{TR}(L,x_0,s)= \frac{\widetilde{q}_N(L,x_0,s)}{1-\widetilde{j}_{L,N}(L,x_0,s)}, \label{surv}
\end{align}
where we have defined $\widetilde{Z}(\cdot,s)=\int_0^\infty~dt~e^{-st}~Z(\cdot,t)$ as the Laplace transform of the function $Z(\cdot,t)$. \eref{surv} can be used to obtain all the moments of the completion times. For instance, the MFPT can be defined as 
\begin{eqnarray}
   \langle \mathcal{T}_N^\text{TR}(L,x_0) \rangle  \equiv \int_0^\infty dt~  \mathcal{Q}_N^\text{TR}(L,x_0,t)= \frac{\langle T_N(L,x_0) \rangle}{\epsilon_0(L,x_0,N)}, 
   \label{MFPT-new}
\end{eqnarray}
where $\langle T_N(L,x_0) \rangle=\int_0^\infty~dt~q_N(L,x_0,t)$ is the mean fastest first passage time out of $N$ searchers to reach either the target or the resetting threshold and 
$\epsilon_0(L,x_0,N)=1-\int_0^\infty~dt~j_{L,N}(L,x_0,t)$ is the splitting probability that any of the $N$ searchers reaches the target first before hitting the threshold.

Since the searchers are non-interacting and identical in nature, we can write $q_N(L,x_0,t) = \left[ Q(L,x_0,t) \right]^N\label{qng}$, where $ Q(L,x_0,t)$ is the survival probability of a single searcher, starting at $x_0 \in [0,L]$ upto time $t$, with absorbing boundaries at both $0$ and $L$ \cite{majumdar2020extreme}. On the other hand, $j_{L,N}(L,x_0,t)$ consists of current to the resetting threshold from a single searcher i.e. $j_{L,1}(L,x_0,t)$ barring the rest $N-1$ of them which survive with probability $ \left[ Q(L,x_0,t) \right]^{N-1}$ so that $j_{L,N}(L,x_0,t)=N j_{L,1}(L,x_0,t) \left[ Q(L,x_0,t) \right]^{N-1}\label{jng}$ (the factor $N$  accounts for the fact that any of the $N$ searchers can contribute to $j_{L,1}(L,x_0,t)$). Substituting these statistical metrics into \eref{MFPT-new}, we arrive at 
\begin{eqnarray}
          \langle \mathcal{T}_N^\text{TR}(L,x_0) \rangle =\frac{\int_0^\infty dt~ [Q(L,x_0,t)]^N}{1-\int_0^\infty dt~N j_{L,1}(L,x_0,t) [Q(L,x_0,t)]^{N-1}}, \label{mfpt}
\end{eqnarray}
which is a general expression for the MFPT and a key result of this paper. 
Importantly, the derivation does not rely on specific dynamics of the searchers or target configurations, and moreover, the framework can be generalized to higher-dimensional search processes.

\emph{\textbf{Ballistic search under TR mechanism.}---} To illustrate the theoretical framework developed above, we examine the paradigmatic case of ballistic searchers with randomized velocity. Such models have been commonly used to describe navigation pattern for some foragers such as microzooplankton, jackals and marine predators \cite{reynolds2009optimising,sims2008scaling}, to scrutinize optimal strategies for identifying randomly located target sites \cite{santos2004optimal,james2008optimizing} and also in statistical physics \cite{villarroel2018continuous,biroli2024exact,martens2012probability,sprenger2023dynamics,bechinger2016active}. Crucially, its analytical tractability offers valuable insights that enhance our understanding of TR-facilitated target search processes. Subsequently, we have extended our analysis to diffusive searchers.

We assume that each ballistic searcher moves with a random initial velocity drawn from a distribution $\phi(v)$. If the target is found \textit{a priori}, the process therein ends. However, if the resetting threshold is encountered prior to the target, all the searchers are simultaneously reset to $x_0$ following which they draw new random velocities from the same distribution. As outlined above, we first derive the survival probability and the probability current for a single ballistic searcher (Sec. S1 in \cite{SI}) 
\begin{align}
    Q(L,x_0,t)&=\Phi\left(\frac{L-x_0}{t}\right) + \Phi\left(\frac{x_0}{t}\right),\\ 
    ~j_{L,1}(L,x_0,t)&=\frac{L-x_0}{t^2}~ \phi\left(\frac{L-x_0}{t}\right), \label{jorv}
\end{align}
where $ \Phi(v)=\int_0^v dv'\phi(v')$; and substitute these into \eref{mfpt} to obtain the MFPT. For symmetric velocity distribution and  $N=1$, this reads \cite{SI}
\begin{align}
    \langle \mathcal{T}_1^\text{TR}(L,x_0) \rangle
    =2L\int_0^{\infty} dv~\frac{\phi(v)}{v}. \label{ball-single}
\end{align}
Remarkably, the MFPT does not depend on the starting point $x_0$ for any
symmetric velocity distribution $\phi(v)$. Notably, the integral in \eref{ball-single} converges only when $\phi(v) \to 0$ as $v\to 0$ so that $\phi(v)/v$ becomes integrable around $v=0$. Simply put, if $\phi(v)$ does not vanish as $v\to 0$, there is a nonzero probability of observing realizations with searchers acquiring an almost vanishing velocity. These searchers will hit either the target or the resetting threshold in an exceedingly large timescale resulting in a divergent MFPT.

\noindent
\textit{MFPT for multiple searchers ($N>1$)}:
To continue the analysis for multiple searchers, we will assume that the velocities are chosen from a probability distribution given by $ \phi(v)=\frac{1}{2 v_0}e^{-|v|/v_0}$. While the MFPT for a single searcher is diverging, as evident from \eref{ball-single}, the MFPT for the collective walkers e.g., $N\ge 2$ turns out to be a finite quantity. For the exponential kernel, the MFPT can be scaled as $ \langle   \overline{\mathcal{T}}_N^\text{TR}(u \equiv \frac{x_0}{L}) \rangle= \frac{v_0}{x_0} \langle \mathcal{T}_N^\text{TR}(L,x_0) \rangle = {\cal F}\left( u,N\right) $, where $\mathcal{F}$ is the scaling function and depends only on $N$ and a dimensionless ratio $0\le u\le 1$. Computation of ${\cal F}(u,N)$ for $N>1$ is a challenging task that we undertake in \cite{SI}, yielding

{\footnotesize
\begin{align}
    \mathcal{F}(u,N)=\frac{\int_0^{\infty} dt\, \left[1- \frac{1}{2}\, e^{-u/t}
-\frac{1}{2}\, e^{-(1-u)/t}\right]^{N}}{N\, u^2\, \int_0^{\infty}
\frac{dt}{2t^2}\, e^{-u/t}\, \left[1- \frac{1}{2}\, e^{-u/t}
-\frac{1}{2}\, e^{-(1-u)/t}\right]^{N-1}}\,.\label{scaling}
\end{align} 
} For $N=2$, a closed form expression can be obtained following a non-trivial computation (Sec. S3 in \cite{SI})
\begin{eqnarray}
 {\cal F}(u,2)= \frac{2}{u\,(3-2u)}\, \left[\ln 2 - u\,\ln u
-(1-u)\, \ln (1-u)\right], 
\end{eqnarray}
while for $N>2$, we can extract the following asymptotic forms which will be useful in later analysis
\begin{eqnarray}
\label{FuN_asymp}
{\cal F}(u,N) \approx \begin{cases}
&\hspace{-0.3cm}\frac{a_N}{2^N-1}\,  \frac{1}{u} \quad\quad\quad\quad\quad\quad\quad\quad
\quad {\rm as}\quad u\to 0\, 
\vspace{0.2cm}
\\
&\hspace{-0.25cm} a_N- N\, (1-u)\, \ln(1-u)\,\quad
{\rm as}\quad u\to 1  ,
\end{cases}
\end{eqnarray} 
where $a_N= \int_0^{\infty} \frac{dx}{x^2}\left(1- e^{-x}\right)^{N}$ is a constant. The scaling function ${\cal F}(u,N)$, evaluated as a function of $u$ for fixed $N$, diverges
as $u\to 0$, and decreases
monotonically with increasing $u$ when $N<4$. Interestingly, for $N \ge 4$, the function exhibits non-monotonic behavior: it first develops a local minimum, followed by a local maximum, and eventually approaches a global minimum as $ u\to 1 $, see \fref{traj}(a) and a representative case for $N=7$ in \fref{traj}(b) (solid curve).

Delving further, we observed that, in the absence of TR ($L\to \infty$ or equivalently
$u=x_0/L \to 0$ limit), the FPT density decays as $\sim \frac{N x_0}{2^N t^2}$, leading to a divergent MFPT for any $N$. In stark contrast, under optimal TR conditions at $u=1$, the distribution exhibits a faster decay, $f_{\mathcal{T}_N^{\text{TR}}}(x_0,t) \sim t^{-N - 1}$, ensuring a finite MFPT as $N \ge 2$ (see Sec. S8 for the derivation of the exact and universal FPT density that holds for arbitrary symmetric velocity distribution).

\textbf{\textit{Optimization and limiting behavior of the MFPT.--}}
To study the optimal behavior, we  rewrite the MFPT by summing over the stochastic timescales 
\begin{eqnarray}
 \langle   \overline{\mathcal{T}}_N^\text{TR}(u) \rangle = \langle {\mathcal{N}}_{\text{TR}} (u,N) \rangle \times \langle \overline{\mathbb{T}}_N^{\text{TR}}(u) \rangle+\langle \overline{\mathbb{T}}_N^{\text{FP}}(u) \rangle,~~ \label{fp-2}
\end{eqnarray}
where $\langle {\mathcal{N}}_{\text{TR}} (u,N) \rangle = \frac{ 1-\epsilon_0 (u,N)}{ \epsilon_0 (u,N)}$ is the mean number of TR events upto the completion, and $ \langle \overline{\mathbb{T}}_N^{\text{TR}}(u) \rangle$ is the mean conditional waiting time between consecutive TR events and $ \langle \overline{\mathbb{T}}_N^{\text{FP}}(u) \rangle$ is the mean time for the first-passage to the target conditioned on \textit{no} TR events. Numerical plots of these quantities are given in \fref{traj}(b) (details in \cite{SI}). 

In the limit $u\to 1$ when the resetting threshold is at the closest proximity to $x_0$, the resetting events are frequent and we find $\langle {\mathcal{N}}_{\text{TR}} (u \to 1,N) \rangle =2^N-1$ \cite{SI}. However, since all such resetting events consume almost no time with $\langle \overline{\mathbb{T}}_N^{\text{TR}}(u \to 1) \rangle \to 0$, the cumulative contribution to the MFPT from TR events becomes almost negligible.  Notably, the finite contribution to the MFPT in this case comes solely from those correlated events where all the ballistic searchers collectively sample negative velocities right after the resetting or at the inception to drift right toward the target, marking a successful completion. Thus, $\langle \overline{\mathbb{T}}_N^{\text{FP}}(u\to 1) \rangle$ effectively contributes to the MFPT whose asymptotic form is given by \eref{FuN_asymp}. The typical trajectories in this limit (see the bottom-right panel in \fref{traj}b) confirm this observation. Quite interestingly, this emergent global optimality at $u \to 1$ aligns with a widely used threshold control strategy in chaos theory \cite{sinha2001using,bhowmick2014targeting}.

However, when the searchers start from a finite distance away from the resetting threshold ($u \lessapprox 1$), 
$\langle \overline{\mathbb{T}}_N^{\text{TR}}(u) \rangle$ become finite (recall that it is exactly zero for the $u \to 1$ limit) and $\langle {\mathcal{N}}_{\text{TR}} (u,N) \rangle$ remains large enough so that the first term in \eref{fp-2} dominates the MFPT 
leading to the local maxima observed in \fref{traj}(b) 
(corresponding trajectories shown in  the top-right figure for $u=0.9$). Further tuning initial configuration of the searchers in the direction of the target (i.e. reducing $u$ further) ensures $\epsilon_0>1/2$ (see Fig. (S2) in \cite{SI}) 
and subsequently, we see a gradual decrease in the MFPT upto 
a local minimum. 
Stochastic realizations at the bottom-left of \fref{traj}(b)  depict this scenario.

Moreover, in the limit $u\to 0$ where the threshold is far away from the starting position, each searcher can acquire a positive or negative velocity with probability $1/2$ for the symmetric velocity distribution and thus, there is a finite probability $\frac{1}{2^N}$ that all the $N$-searchers possess a positive velocity that is away from the target. The probability of selecting a unfavorable positive velocity, albeit vanishingly small, remains finite and yield in MFPT with a diverging form $\sim 1/u$ given by \eref{FuN_asymp} (Sec. S5 in \cite{SI}).  Contributing realizations (for instance, shown in the top-left panel in \fref{traj}b for $u=0.0015$) corroborate this observation.

Finally, examining the MFPT with respect to $N$, we find that depending on $u$ and the interaction mediated by $N$, this variation can be either monotonic or non-monotonic. Nevertheless, for all $N$, the MFPT attains its minimum at $u = 1$ (see Fig. S2), further supporting our claim in the ballistic search scenario.

\begin{figure}
    \centering
    \includegraphics[width=7cm]{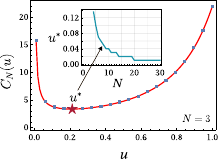}
    \caption{Behavior of the cost function with respect to $u=x_0/L$ for $N=3$ and $\beta=1$. The cost attains a minimum at an optimal $u^*$ -- the variation of which as a function of $N$ is shown in the inset panel. The squares represent data from simulations. }
    \label{fig-cost}
\end{figure}

\emph{\textbf{Cost of TR mechanism.}---} A central quest in the resetting discipline is to harness the cost, be dynamic \cite{de2020optimization,sunil2023cost,sunil2024minimizing} or thermodynamic \cite{tal2020experimental,tal2025smart,olsen2024thermodynamic,gupta2020work,pal2023thermodynamic}, of the overall
process. Imitating the penalty during a power grid breakdown \cite{ajjarapu1994optimal,feng2000comprehensive}, we posit, for each resetting event, a cost that is proportional to the MFPT as well as the total number of TR events. Henceforth, the goal is
to determine the optimal operation of this system, for which
the cost function
\begin{align}
    C_N (u)=\langle \overline{\mathcal{T}}_N^\text{TR}(u) \rangle + \beta N  \langle {\mathcal{N}}_{\text{TR}} (u,N) \rangle,  \label{cost}
\end{align}
is minimal. Here, $\beta$ is a constant that can be thought of cost per resetting event  for a single particle and thus, we multiply by $N$ to enumerate the total cost for the ``reset operation''. For the ballistic searchers, \fref{fig-cost} shows that $ C_N (u)$ attains a minimum for an optimal $u=u^*$ for a fixed $N$. Note that unlike the MFPT, the cost function is not globally minimized at $u\to 1$ since  in this limit the resetting frequency is significantly high and the second term in \eref{cost} dominates. On the other hand, if the resetting threshold is far away from the starting position, the MFPT gradually increases, contributing to diverging cost.  
In the inset of \fref{fig-cost}, we show the variation of $u^*$ with respect to $N$, further noting that large number of searchers can systematically reduce the optimal distance $u^* \to 0$. Interestingly, the overall cost function remains optimized despite the increase in individual cost through $\beta$, accompanied by a gradual rise in $u^*$ \cite{SI}. Furthermore, $u^*$ decreases with increasing $\beta$ for a fixed $N$, thus reflecting the high cost of frequent resets \cite{SI}.
Evidently, this collective cost optimization is different than that of the MFPT, indicating that a judicious choice of optimal $u$ can leverage to a decision based biasing between the MFPT and the cost function.

\emph{\textbf{Conclusions.}---} 
Motivated by threshold-driven phenomena that refer to processes or events that occur only when a system surpasses a critical threshold, we study the first passage properties of an arbitrary search dynamics under threshold resetting mechanism. We develop a general theoretical framework and study the optimization properties of the MFPT which is a useful metric in learning the efficiency of a target search process. The optimization reported here also resonates with the performance maximization by the agents (robots or foraging animals) that collaborate and achieve tasks (for instance finding the desired target between many competitive ones by a fish school in \cite{couzin2011uninformed}) in an optimal way.

In a classical foraging model where the searchers move ballistically, we discover a non-trivial optimization that suggests that persistent resetting to the threshold can render the MFPT globally minimum -- an effect reminiscent to the threshold mechanism in nonlinear chaotic systems \cite{sinha2001using,bhowmick2014targeting}. Extending the analysis to $N$ diffusive searchers also reveals rich optimization behaviors of the MFPT \cite{SI,diffusive_TR}. By defining a cost function for such systems, we further demonstrate that the collective TR mechanism can achieve optimal cost. 

 To extend the formalism to capture more realistic scenarios, we can incorporate an overhead time for information transfer. This can be done in two ways: First, we can add a random (or deterministic) information passing time $\mathcal{T}^O$ whenever a searcher hits the threshold, after which all of them are reset to the initial position. In this case, one can show that the overall MFPT of the system would increase 
(however, this computation excludes the time required to transfer information to the others once a searcher successfully locates the target) \cite{diffusive_TR}. A more involved extension would be to consider space–time coupled return protocols, where after a TR event searchers return to the initial configuration via finite-time spatial dynamics \cite{pal_search_2020,tal2020experimental,biswas2024search}.

Finally, it is worth noting that the formalism can also be applied to extract higher order moments of the search time in higher dimensions and to the dynamics of non-coordinate variables
like a fluctuating potential as in the case of firing neurons where the membrane potential of an individual synapse can reset after reaching a critical value \cite{burkitt2006review,taillefumier2013phase}. Ongoing works also include characterizing nonequilibrium stationary states for such long-range systems under TR \cite{diffusive_TR,BMS2025}. Given its simple implementation yet achieving target functionality makes the TR mechanism a promising avenue in the target search problem which can have far reaching implications beyond statistical physics to foraging, control theory, finance, structural failure-repair analysis and in computing systems.

\emph{\textbf{Acknowledgments.}---} SNM thanks M. Biroli, S. Redner and G. Schehr for discussions on related models. AP sincerely thanks D. Ghosh for pointing towards some relevant works. The numerical calculations reported in
this work were carried out on the Kamet cluster, which is maintained and supported by the Institute of Mathematical Science’s High-Performance Computing Center. AB and AP gratefully acknowledge research support from the Department of Atomic Energy, Government of India. SNM acknowledges support from ANR Grant No. ANR- 23- CE30-0020-01 EDIPS. 
 SNM and AP thank the Higgs Center for Theoretical Physics, Edinburgh, for hospitality during the workshop ``New Vistas in Stochastic Resetting'' and Korea Institute for advanced Study(KIAS), Seoul, for hospitality during the conference ``Nonequilibrium Statistical Physics of Complex Systems''  where several discussions related to the project took place. SNM and AP also acknowledge the research support from the International Research Project (IRP) titled ``Classical and quantum dynamics in out of equilibrium systems'' by CNRS, France.

\bibliography{fpusr}

\section*{End Matter}
\textit{Diffusive searchers under TR.} 
Let us consider the case of $N$ diffusive searchers with the same set-up as for the ballistic walkers. To proceed, we will need the single searcher diffusive propagator $G(x,t|x_0)$ in the presence of two absorbing boundaries at $x=0$ and $x=L$. This is well known in literature and is given by \cite{redner2001}
\begin{align}
    G(x,t|x_0)=\frac{2}{L} \sum _{n=0}^{\infty}  \sin \left(\frac{\pi  n x}{L}\right) \sin \left(\frac{\pi  n x_0}{L}\right) e^{-\frac{n^2 \pi ^2 D t}{L^2}}, 
\end{align}
where $D$ is the diffusion constant.
Using this, we can calculate the survival probability and the currents through the threshold and the target in terms of two quantities $u$ and $\tau_d=L^2/D$ so that \cite{SI}
\begin{align}
    &Q(u,t)=\frac{2}{\pi}\sum _{n=1}^{\infty} \left(\frac{1-(-1)^n }{  n} \right)\sin \left(n\pi u\right) e^{-\frac{n^2 \pi ^2 t}{\tau_d}},
   \\
    & j_{0,1}(u,t)=\frac{2 \pi }{\tau_d} \sum _{n=1}^{\infty} n \sin  \left(n\pi u\right) e^{-\frac{n^2 \pi ^2  t}{\tau_d}},\label{jond}\\
  & j_{L,1}(u,t) =-\frac{2 \pi}{\tau_d} \sum _{n=1}^{\infty} n (-1)^n \sin  \left(n\pi u\right) e^{-\frac{n^2 \pi ^2 t}{\tau_d}}, 
\end{align}
\begin{figure}[H]
    \centering
    \includegraphics[width=7.5cm]{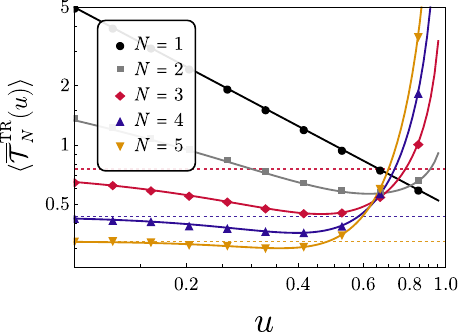}
    \caption{Diffusive searchers under TR: Mean first-passage time (MFPT) as a function of the threshold parameter $u$ for different values of $N$ as in \eref{mfpt-diff}. The horizontal dashed lines (for $N \geq 3$) represent the MFPT in the absence of threshold resetting, corresponding to the $u \to 0$ limit (note that the MFPTs for $N=1,2$ in this limit are diverging). The fact that the MFPT curves fall below these baselines highlights the enhanced search efficiency achieved through the TR mechanism, also in the case of diffusive searchers. The markers represent the simulation results.}
    \label{fig-diff}
\end{figure}

The MFPT can be written in a dimensionless form as $\langle \overline{\mathcal{T}}_N^{\text{TR}}(u) \rangle=\langle \mathcal{T}_N^{\text{TR}}(x_0,L) \rangle  /\tau_d =\mathcal{F}(u,N)$. Skipping details from \cite{SI}, we obtain an exact expression of the scaling function $\mathcal{F}(u,N)$ representing the MFPT which is given by
\begin{widetext}
\begin{align}
\mathcal{F}(u,N)
&=\frac{1}{u^2}\left(\frac{ \int_0^\infty dt~ \left[\frac{2}{\pi}\sum _{n=1}^{\infty} \left(\frac{1-(-1)^n }{  n} \right)\sin \left(n\pi u\right) e^{-n^2 \pi ^2 t}\right]^N}{N\int_0^\infty dt~ \left[2 \pi  \sum _{n=1}^{\infty} n \sin  \left(n\pi u\right) e^{-n^2 \pi ^2  t}\right] \left[\frac{2}{\pi}\sum _{n=1}^{\infty} \left(\frac{1-(-1)^n }{  n} \right)\sin \left(n\pi u\right) e^{-n^2 \pi ^2 t}\right]^{N-1}}\right),\label{mfpt-diff}
\end{align}
\end{widetext}
which we plot in \fref{fig-diff} as a function of $u$ for different $N$. Summarizing the results from the plot and subsequent analysis \cite{SI}, we note that for a single diffusive searcher, the MFPT decreases as $u^{-1}$. It is the lowest when $L\to x_0$ ($u\to 1$) where $ \langle \overline{\mathcal{T}}_1^{\text{TR}}(u=1) \rangle \rangle =1/2$. However, for $N \ge 2$, the MFPT exhibits distinct optimization features. It should be noted that for $N=1$ and $N=2$, the TR mechanism always reduces the search time. It is observed from \fref{fig-diff} that TR can facilitate the search process while compared to the same in the absence of TR as for intermediate values of $u$, the MFPT curves go below their respective dashed lines—each representing the reset-free case for different $N$—clearly indicating that threshold resetting enhances search efficiency. Although it is known that the MFPT for diffusive searchers is finite only for $N \geq 3$ (see \cite{lindenberg1980lattice,krapivsky2010maximum,SI}), the introduction of the TR mechanism enables a further reduction by optimally constraining the searchers' trajectories, leading to an optimal resetting threshold in the range $0 < u < 1$ for all $N \geq 2$. This optimization observed in the diffusive case, alongside similar results for ballistic searchers, demonstrates that TR is a robust, event-driven mechanism capable of significantly accelerating target search in complex collective systems.


\begin{titlepage}
\title{Supplemental material for \\ \underline{``Target search optimization by threshold resetting"}}
\maketitle
\end{titlepage}

\onecolumngrid
\setcounter{page}{1}
\renewcommand{\thepage}{S\arabic{page}}
\setcounter{equation}{0}
\renewcommand{\theequation}{S\arabic{equation}}
\setcounter{figure}{0}
\renewcommand{\thefigure}{S\arabic{figure}}
\setcounter{section}{0}
\renewcommand{\thesection}{S\arabic{section}}
\setcounter{table}{0}
\renewcommand{\thetable}{S\arabic{table}}

This Supplemental Material provides detailed derivations of some of our main results highlighted in the Letter. Moreover, it also provides additional discussions that support our finding as announced in the Letter. 


\tableofcontents

\section{Survival probability and probability current of a single ballistic searcher}
In this section, we compute the survival probability $Q(L,x_0,t)$ and the current through the resetting threshold $j_{L,1}(L,x_0,t)$ for a single ballistic searcher as used in Eq. (5) and (6) of the main text. If $v_i$ denotes the random initial velocity of the $i^{th}$ searcher then it can survive up to time $t$ without hitting either target or resetting threshold in two ways: First, if $v_i t< L-x_0$ when $v_i>0$ so that the walker starts moving towards the resetting threshold but failing to reach the same in the given time and second, when $|v_i| t <x_0$ if the initial velocity $v_i<0$ so that the walker starts moving towards the target yet fails to hit it by time $t$. 
Summing over these two possibilities, we can write the survival probability as
\begin{align}
    &Q(L,x_0,t)=\int_0^\infty dv~\phi(v)~ \theta(L-x_0 - v t) + \int_{-\infty}^0 dv~\phi(v)~ \theta(x_0 - |v| t), \label{surv-ball}
\end{align}
where $\theta(z)$ is the Heaviside theta function which takes values unity only when $z>0$, and $0$ otherwise. Assuming $\phi (v)$ to be symmetric so that $\phi(-v)=\phi(v)$, one can rewrite the above equation as 
\begin{align}
    Q(L,x_0,t)=\int_0^{\frac{L-x_0}{t}}dv~\phi(v) + \int_{0}^{\frac{x_0}{t}} dv~\phi(v).
\end{align}
Denoting the function $\Phi(v)$ as 
\begin{align}
    \Phi(v)=\int_0^v dv'\phi(v'), \label{cmlv}
\end{align}
we finally arrive at
\begin{align}
      Q(L,x_0,t)=\Phi\left(\frac{L-x_0}{t}\right) + \Phi\left(\frac{x_0}{t}\right), \label{qrv}
\end{align}
which is Eq. (5) of the main text. The probability current through the resetting threshold at $x=L$ is essentially the total number of accumulated particles there at time $t$ so that
\begin{align}
    j_{L,1}(L,x_0,t)&=\int_{0}^\infty dv~\phi(v)~\delta\left(t - \frac{L-x_0}{v}\right)\nonumber\\
    &=\frac{L-x_0}{t^2}~ \phi\left(\frac{L-x_0}{t}\right), \label{jrv}
\end{align}
which is Eq. (6) of the main text. Here, $\delta(x)$ is the Dirac-delta function. Note that only the positive velocities contribute to the current through the threshold  while the negative velocities contribute to the current through the target at the origin
given by
\begin{align}
    j_{0,1}(L,x_0,t)&=\int_{-\infty}^0 dv~\phi(v)~\delta\left(t - \frac{x_0}{v}\right)\nonumber\\
    &=\frac{x_0}{t^2}~ \phi\left(\frac{x_0}{t}\right). 
\end{align}

\section{MFPT of a single ballistic searcher under TR Mechanism Eq. (7)}
This section extends the derivation of the MFPT of a single searcher under TR mechanism which was announced in Eq. (7) of the main text.
To this end, we start by deriving the numerator in Eq. (3) of the main text by setting $N=1$ which yields the unconditional MFPT for a single particle to reach either the resetting threshold or the target 
\begin{align}
    \langle T_{1}(L,x_0) \rangle &= \int_0^\infty dt~ Q(L,x_0,t) \nonumber\\
    &=\int_0^\infty dt \int_0^\infty dv~\phi(v)~ \theta(L-x_0 - v t)  +\int_0^\infty dt \int_{-\infty}^0 dv~\phi(v)~ \theta(x_0 - |v| t) \nonumber\\
    &= \int_0^\infty dv~\phi(v)\int_0^\infty dt~ \theta(L-x_0 - v t) + \int_{-\infty}^0 dv~\phi(v)~\int_0^\infty dt \theta(x_0 - |v| t) \nonumber\\
    &= \int_0^{\infty}dv~\phi(v)~ \left(\frac{L-x_0}{v}\right) +\int_{-\infty}^0 dv~\phi(v)~ \left(\frac{x_0}{|v|}\right).
\end{align}
Using the symmetry property of the velocity distribution i.e.  $\phi(-v)=\phi(v)$, we can further simplify the above equation to find
\begin{align}
     \langle T_{1}(L,x_0) \rangle&= \int_0^{\infty} dv~\phi(v)~ \left(\frac{L-x_0}{v}\right) +\int_0^{\infty} dv~\phi(v)~ \left(\frac{x_0}{v}\right) \nonumber\\
    &=L\int_0^{\infty} dv~\frac{\phi(v)}{v}.
\end{align}

Let us turn our attention to compute the denominator of Eq. (3) from the main text for $N=1$. We first note that it is just the splitting probability $\epsilon_0(L,x_0,1)$ of a single searcher to reach the target located at the origin before hitting the threshold. For a symmetric velocity distribution, $\epsilon_0(L,x_0,1)$ is simply $1/2$ since any searcher can acquire a positive (negative) velocity with probability $1/2$ indifferent to the initial coordinate $x_0$. This can also be seen formally as 
\begin{align}
    \epsilon_0(L,x_0,1)&=1-\int_0^\infty dt~ j_{L,1}(L,x_0,t) \nonumber\\
    &=\int_0^\infty dt~ j_{0,1}(L,x_0,t) \nonumber\\
    &=\int_0^\infty dt ~\frac{x_0}{t^2}~ \phi(x_0/t)\nonumber\\
    &=\frac{1}{2}.
\end{align}
Putting together both the numerator and denominator in Eq. (3) of the main text, the MFPT under TR mechanism for a single searcher takes the form
\begin{align}
    \langle \mathcal{T}_1^\text{TR}(L,x_0) \rangle=\frac{ \langle T_{1}(L,x_0) \rangle }{ \epsilon_0(L,x_0,1)}=2L\int_0^{\infty} dv~\frac{\phi(v)}{v},
\end{align}
which is Eq. (7) of the main text.

\section{Derivation of the scaling function ${\cal F}(u,N)$ as in Eq. (8)}
\begin{figure}
    \centering
    \includegraphics[width=8cm]{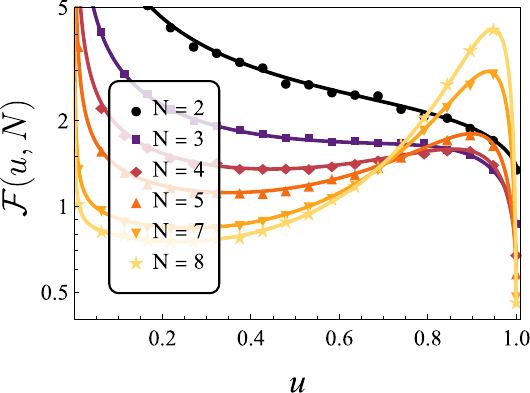}
    \caption{Variation of the scaling function $\mathcal{F}(u,N)$ as in \eref{Fu_exact.1} with respect to $u$ for different choices of $N$. The markers represents the results from numerical simulation which match perfectly with the theoretical curves indicated by the solid lines.}
    \label{tvsu}
\end{figure}

In this section, we derive an exact form of the scaling function ${\cal F}(u,N)$ which was announced in the main text. 
To this end, we use Eq. (4) from the main text and note that we need the single particle survival probability (as in \eref{qrv}) and the probability current to the resetting threshold (as in \eref{jrv}) for the exponential velocity distribution $ \phi(v)=\frac{1}{2 v_0}e^{-|v|/v_0}$. 
In this case, we have $\Phi(v)=\frac{1}{2 } \left(1-e^{-v/v_0}\right)$.
Defining the  following variables 
\begin{align}
    u=x_0/L, ~~ \tau_b=L/v_0, \label{utau}
\end{align}
where $\tau_b$ is the timescale of a ballistic walker with speed $v_0$ and using \eref{qrv} and \eref{jrv}, we can now compute the survival probability and the currents to the resetting threshold and the target respectively which take the following form
\begin{align} 
    & Q(u,t)=1-\frac{1}{2} \exp \left({-\frac{\tau_b(1-u) }{t}}\right)-\frac{1}{2}\exp\left({-\frac{\tau_b u}{t}}\right), \label{qu} \\
    & j_{L,1}(u,t)=\frac{\tau_b (1-u)}{2 t^2}\exp \left({-\frac{\tau_b(1-u) }{t}}\right), \label{jlu}\\
    & j_{0,1}(u,t)=\frac{\tau_bu}{2 t^2}\exp \left({-\frac{\tau_b u}{t}}\right). \label{jou}
    \end{align}
We now recall from the main text that the splitting probability for any of the $N$ particles to hit the target is given by
\begin{align}
    \epsilon_0(L,x_0,N)&=1-\int_0^\infty dt~ j_{L,N}(L,x_0,t) \nonumber \\
    &=1-\int_0^\infty dt~N j_{L,1}(L,x_0,t) [Q(L,x_0,t)]^{N-1} \nonumber\\
    &=\int_0^\infty dt~N j_{0,1}(L,x_0,t) [Q(L,x_0,t)]^{N-1}. \label{exit-o}
\end{align}
Plugging the results from \eref{qu}-(\ref{exit-o}) into Eq. (4) of the main text, we find
\begin{align}
       \langle\mathcal{T}_N^\text{TR}(L,x_0) \rangle=\frac{\int_0^\infty dt~ \left[1-\frac{1}{2} \exp \left({-\frac{[1-u]\tau_b }{t}}\right)-\frac{1}{2}\exp\left({-\frac{\tau_b u}{t}}\right)\right]^N}{\int_0^\infty dt~N \left[\frac{\tau_bu}{2 t^2}\exp \left({-\frac{u\tau_b }{t}}\right)\right] \left[1-\frac{1}{2} \exp \left({-\frac{[1-u]\tau_b }{t}}\right)-\frac{1}{2}\exp\left({-\frac{\tau_b u}{t}}\right)\right]^{N-1}}.
\end{align}
Now we make the change of variable as $t \to \frac{\tau_b}{t}$ to obtain
\begin{align}
     \langle\mathcal{T}_N^\text{TR}(L,x_0) \rangle= \frac{\tau_b\int_0^\infty dt \left[1-\frac{1}{2} \exp \left({-\frac{1-u }{t}}\right)-\frac{1}{2}\exp\left({-\frac{ u}{t}}\right)\right]^N}{N u \int_0^\infty dt \left[\frac{1}{2 t^2}\exp \left({-\frac{u }{t}}\right)\right] \left[1-\frac{1}{2} \exp \left({-\frac{1-u }{t}}\right)-\frac{1}{2}\exp\left({-\frac{u}{t}}\right)\right]^{N-1} }.
\end{align}
Noting $\tau_b=\frac{1}{u} \frac{x_0}{v}$ from \eref{utau} and substituting into the above expression for MFPT, we find
\begin{align}
      &\langle\mathcal{T}_N^\text{TR}(L,x_0) \rangle = \frac{\frac{x_0}{v}\int_0^\infty dt \left[1-\frac{1}{2} \exp \left({-\frac{1-u }{t}}\right)-\frac{1}{2}\exp\left({-\frac{ u}{t}}\right)\right]^N}{N u^2 \int_0^\infty dt \left[\frac{1}{2 t^2}\exp \left({-\frac{u }{t}}\right)\right] \left[1-\frac{1}{2} \exp \left({-\frac{1-u }{t}}\right)-\frac{1}{2}\exp\left({-\frac{u}{t}}\right)\right]^{N-1} }, \nonumber\\
      \implies &\mathcal{F}(u,N)=\frac{v_0}{x_0}\langle\mathcal{T}_N^\text{TR}(L,x_0) \rangle=\frac{\int_0^\infty dt \left[1-\frac{1}{2} \exp \left({-\frac{1-u }{t}}\right)-\frac{1}{2}\exp\left({-\frac{ u}{t}}\right)\right]^N}{N u^2 \int_0^\infty dt \left[\frac{1}{2 t^2}\exp \left({-\frac{u }{t}}\right)\right] \left[1-\frac{1}{2} \exp \left({-\frac{1-u }{t}}\right)-\frac{1}{2}\exp\left({-\frac{u}{t}}\right)\right]^{N-1} }, \label{Fu_exact.1}
\end{align}
which is essentially Eq. (8) of the main text. In Fig. (\ref{tvsu}), we show the variation of the scaling function with $u$ for different $N$. The scaling function ${\cal F}(u,N)$, evaluated as a function of $u$ for fixed $N$, diverges
as $u\to 0$, and decreases
monotonically with increasing $u$ when $N<4$. Interestingly, for $N \ge 4$, the function exhibits non-monotonic behavior: it first develops a local minimum, followed by a local maximum, and eventually approaches a global minimum as $ u= 1 $. 

\begin{figure}
    \centering
    \includegraphics[width=8cm]{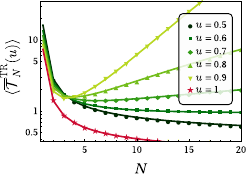}
    \caption{Variation of the MFPT of ballistic searchers under TR with respect to $N$ for various values of $u$. Note that for $u=1$ the MFPT is lowest for any values of $N$ as shown by the red line. The markers represents results from numerical simulation.}
    \label{tvsn}
\end{figure}

In Fig. (\ref{tvsn}) we also show the variation of the MFPT with respect to $N$ for several choices of $u$. From the figure it is evident that (i) the MFPT for any $u<1$ can be monotonic or non-monotonic in nature with respect to $N$, however (ii) the envelope for MFPT corresponding to $u=1$ remains the lowest for any choice of $N$. The results reconcile with the key result for the ballistic searchers that $u=1$ is the optimal value for the collective ballistic search. It also implies that the long-range interaction reinforced by the collective resetting of $N$ searchers works in the most optimal way when $u=1$.

\section{Derivation of the closed form for the scaling function $\mathcal{F}(u,N)$ for $N=2$ as in Eq. (9)}
Here we derive the scaling function (or the scaled MFPT) for $N=2$ ballistic searchers i.e. $\mathcal{F}(u,2)$. It is convenient to make the change of variable $t=1/x$ and rewrite the scaling function (Eq.  (8) of the main text or \eref{Fu_exact.1}) as
\begin{equation}
{\cal F}(u,N)=\frac{2}{N\, u^2}\, \frac{I_N(u)}{J_N(u)}\, ,
\label{Fu_exact.2}
\end{equation}
where $I_N(u)$ and $J_N(u)$ are given by
\begin{eqnarray}
I_N(u)&= & \int_0^{\infty} \frac{dx}{x^2}\, 
\left[1- \frac{1}{2}\, e^{-u\, x}-\frac{1}{2}\, e^{-(1-u)x}\right]^{N} \label{Inu.1} \\
J_N(u) &=& 
\int_0^{\infty} dx\, e^{-u\, x}\, \left[1- \frac{1}{2}\, 
e^{-u\, x}-\frac{1}{2}\, e^{-(1-u)x}\right]^{N-1}\, .
\label{Jnu.1}
\end{eqnarray}
For general $N\ge 2$, it is not easy to compute the two integrals $I_N(u)$ and $J_N(u)$, and
hence ${\cal F}(u,N)$ in Eq. (\ref{Fu_exact.2}). However for $N=2$, as we show below,
one can compute ${\cal F}(u,2)$ exactly for all $u\in [0,1]$. 
This will serve also as a benchmark to compare
to the asymptotic behaviors of ${\cal F}(u,N)$ for general $N$ in the
two asymptotic limits $u\to 0$ and $u\to 1$.

For $N=2$, let us start with the integral $J_2(u)$ in Eq. (\ref{Jnu.1}) which is relatively
simpler to evaluate. Setting $N=2$ in Eq. (\ref{Jnu.1}) we get
\begin{equation}
J_2(u)=\int_0^{\infty} dx\, e^{-ux}\, \left[1- \frac{1}{2}\, 
e^{-u\, x}-\frac{1}{2}\, e^{-(1-u)x}\right]= \frac{3}{4u}-\frac{1}{2}\, .
\label{J2u.1}
\end{equation}
The computation of $I_2(u)$ in Eq. (\ref{Inu.1}) turns out to be tricky as we show in below. Following
Eq. (\ref{Inu.1}), we write
\begin{equation}
I_2(u)= \int_0^{\infty} \frac{dx}{x^2} \left[ 1- \frac{1}{2}\, e^{-u\, x}- \frac{1}{2}\,
e^{-(1-u)\, x}\right]^2\, .
\label{I2u.1}
\end{equation}
To get rid of the singularity $x^{-2}$ in the integrand, it is convenient to first take
a derivative with respect to $u$. This gives
\begin{equation}
I_2'(u)= \frac{dI_2(u)}{du}=\int_0^{\infty} \frac{dx}{x} 
\left[ 1- \frac{1}{2}\, e^{-u\, x}- \frac{1}{2}\, e^{-(1-u)\, x}\right]\, 
\left[ e^{-u\, x}- e^{-(1-u)\, x}\right]\, .
\label{I2u.2}
\end{equation}
We note one important fact from this equation that will be useful later, namely
\begin{equation}
I_2'\left(u=\frac{1}{2}\right)=0\, .
\label{bc.1}
\end{equation}
The integrand in Eq. (\ref{I2u.2}) still has an  singularity $1/x$. To get rid of this 
term, let us take one more derivative with respect to $u$. Further simplification leads to the following integral that can be trivially computed to yield
\begin{equation}
I_2''(u)= \frac{d^2I_2(u)}{du^2}= \int_0^{\infty} dx\, 
\left[e^{-2 u x}- e^{-2(1-u)x} - e^{-u x}- e^{-(1-u) x}\right]= -\frac{1}{2u}- \frac{1}{2(1-u)}\, .
\label{I2u.3}
\end{equation}
Now, integrating it back with respect to $u$ gives
\begin{equation}
I_2'(u)= -\frac{1}{2} \ln u + \frac{1}{2} \ln (1-u) +C_1\, ,
\label{I2u.4}
\end{equation}
where the unknown constant $C_1=0$ is fixed from the condition in Eq. (\ref{bc.1}). Thus we get
\begin{equation}
I_2'(u)= -\frac{1}{2} \ln u + \frac{1}{2} \ln (1-u)\, .
\label{I2u.5}
\end{equation}
\begin{figure}
    \centering
    \includegraphics[width=8cm]{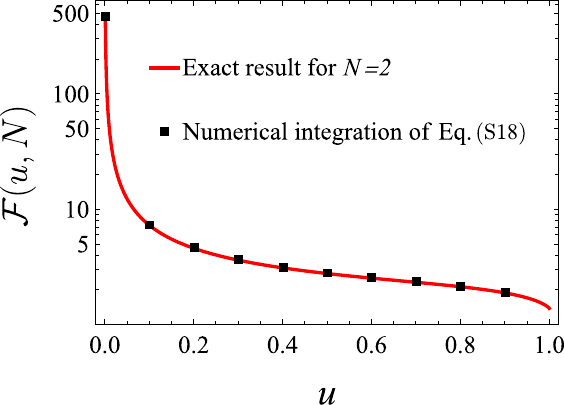}
    \caption{Comparison of the scaling function for $N=2$ as in \eref{F2u_exact} with that obtained via numerical integration of \eref{Fu_exact.1} .}
    \label{N2}
\end{figure}
Integrating once more we get
\begin{equation}
I_2(u)= -\frac{1}{2} \left[u \ln u-u\right] + \frac{1}{2} \left[1-u - (1-u)\ln (1-u)\right]+C_0\, ,
\label{I2u.6}
\end{equation}
where the constant $C_0$ is fixed from the value $I_2(0)$ in
Eq. (\ref{I2u.1}). Thus we finally get
\begin{equation}
I_2(u)= I_2(0) - \frac{1}{2}\, u\, \ln u- \frac{1}{2}\, (1-u)\, \ln(1-u)\, ,
\label{I2u.7}
\end{equation}
where, using Eq. (\ref{I2u.1}),
\begin{equation}
I_2(0)= \frac{1}{4} \, a_2 \, \quad {\rm with}\quad a_2= \int_0^{\infty} \frac{dx}{x^2} \left[1- e^{-u x}\right]^2\, .
\label{I20.1}
\end{equation}
To evaluate the integral $a_2$, we do one integration by part that gives
\begin{equation}
a_2= 2 \int_0^{\infty} \frac{dx}{x}\, e^{-x} \left(1- e^{-x}\right)\, .
\label{a2.1}
\end{equation}
To do the remaining integral in Eq. (\ref{a2.1}),
let us introduce an auxiliary parameter $b$ and define
\begin{equation}
a_2(b)= 2\, \int_0^{\infty} \frac{dx}{x}\, e^{-x} \left(1- e^{-b\, x}\right)\, ,
\label{a2b.1}
\end{equation}
such that $a_2(b=1)= a_2$. Differentiating Eq. (\ref{a2b.1}) with respect to $b$ we get
\begin{equation}
a_2'(b)= \frac{da_2(b)}{db}= 2\, \int_0^{\infty} dx\, e^{-(b+1)\, x}= 
\frac{2}{b+1}\, .
\label{a2b.2}
\end{equation}
Integrating it back with respect to $b$ and using $a_2(b=0)=0$, we get
\begin{equation}
a_2(b)= 2\, \ln (b+1)\, .
\label{a2b.3}
\end{equation}
Consequently,
\begin{equation}
a_2= a_2(b=1)= 2\, \ln 2\, .
\label{a2_final}
\end{equation}
Using this result in Eq. (\ref{I20.1}) gives
\begin{equation}
I_2(0)= \frac{a_2}{4}= \frac{1}{2}\, \ln 2 \, .
\label{I20.2}
\end{equation}
Following Eq. (\ref{I2u.7}), we get the complete expression for $I_2(u)$ as 
\begin{equation}
I_2(u) = \frac{1}{2}\,\left[\ln 2  - u\, \ln u- 
(1-u)\, \ln(1-u)\right]\, . 
\label{I2u.8} 
\end{equation}
Substituting $J_2(u)$ from Eq. (\ref{J2u.1}) and $I_2(u)$ from Eq.  (\ref{I2u.8})
into Eq. (\ref{Fu_exact.2}) with $N=2$ gives the exact scaling function ${\cal F}(u,2)$ as 
\begin{equation}
{\cal F}(u,2)= \frac{2}{u\,(3-2u)}\, \left[\ln 2 - u\,\ln u
-(1-u)\, \ln (1-u)\right]\, , 
\label{F2u_exact}
\end{equation}
as mentioned in Eq. (9) of the main text.  In Fig. (\ref{N2}) we show that the exact result as in \eref{F2u_exact} matches exactly with the numerical integration of \eref{Fu_exact.1} for $N=2$. The limiting behaviors of ${\cal F}(u,2)$ are given by
\begin{eqnarray}
\label{F2u_asymp}
{\cal F}(u,2) \approx \begin{cases}
\frac{2\ln 2}{3\, u} \, \quad\quad\quad\quad\quad\quad\quad\quad\quad\quad\quad 
{\rm as}\quad u\to 0 \\
\\
2\, \ln 2 - 2\ (1-u)\, \ln(1-u) \,\,\, \quad {\rm as}\quad u\to 1\, .
\end{cases}
\end{eqnarray} 
Derivation of the scaling function for general $N$ is subject matter of the next section.

\section{Asymptotic behaviors of ${\cal F}(u,N)$  
for general $N\ge 2$ as in Eq. (10)}

We now consider the scaling function ${\cal F}(u,N)$ in Eq. (\ref{Fu_exact.2})
for general $N\ge 2$. 
Unlike the $N=2$ case, it turns out to be much harder to compute the scaling
function explicitly for general $N>2$. However, as mentioned earlier, 
one can extract
the asymptotic behaviors of ${\cal F}(u,N)$, for arbitrary $N\ge 2$,
in the two limits
$u\to 0$ and $u\to 1$ as shown below.

\subsection{The limit $u\to 0$}

We will show that for general $N\ge 2$, the scaling function ${\cal F}(u,N)$
diverges as $u\to 0$, and this leading divergence originates 
from the denominator $J_N(u)$ in Eq. (\ref{Fu_exact.2}) as $u\to 0$. Hence,
for the numerator $I_N(u)$ in Eq. (\ref{Fu_exact.2}), it is enough only the
leading $u=0$ behavior where it turns out to be finite for all $N\ge 1$.
This becomes evident by setting $u=0$ in Eq. (\ref{Inu.1}), and it gives
\begin{equation}
I_N(0)= \frac{a_N}{2^N}\, \quad {\rm where}\quad a_N= \int_0^{\infty} \frac{dx}{x^2}\left(1- e^{-x}\right)^{N}\, .
\label{In0.1}
\end{equation}
We then need to evaluate $J_N(u)$ in Eq. (\ref{Jnu.1}) in the limit $u\to 0$.
By naively putting $u=0$ in Eq. (\ref{Jnu.1}) one finds that the resulting
integral diverges. To extract the leading divergence of $J_N(u)$ as $u\to 0$,
we first make a change of variable $u\, x=y$ in the integral in Eq. (\ref{Jnu.1}).
This gives
\begin{equation}
J_N(u)= \frac{1}{u}\, \int_0^{\infty} dy\, 
e^{-y}\, \left[1- \frac{1}{2}\, e^{-y}- 
\frac{1}{2}\, e^{-\frac{(1-u)}{u}\, y}\right]^{N-1}\, .
\label{Jnu.2}
\end{equation}
Now, in the limit $u\to 0$, we can safely neglect the third term inside the
parenthesis $[\ldots]$ since the resulting integral is convergent. This gives the
leading order behavior as $u\to 0$
\begin{equation}
J_N(u) \approx \frac{1}{u}\, 
\int_0^{\infty} dy\, e^{-y}\, \left[1-\frac{1}{2}\, e^{-y}\right]^{N-1}\, .
\label{Jnu.3}
\end{equation}
Making the change of variable $z=1- \frac{1}{2}\, e^{-y}$, the integral
can be done exactly leading to
\begin{equation}
J_N(u)\approx \left[\frac{2}{N}\left(1- 2^{-N}\right)\right]\, \frac{1}{u}\, 
\quad {\rm as}\quad u\to 0\, .
\label{Jnu.4}
\end{equation}
Substituting Eq. (\ref{Jnu.4}) and Eq. (\ref{In0.1}) into the exact expression
for ${\cal F}(u,N)$ in Eq. (\ref{Fu_exact.2}) then gives the leading
small $u$ behavior
\begin{equation}
{\cal F}(u,N) \approx \frac{a_N}{2^N-1}\, \frac{1}{u} \quad 
{\rm as}\quad u\to 0\, ,
\label{Fu0.1}
\end{equation}
where $a_N$ is defined in Eq. (\ref{In0.1}). Note that this result is
valid for any $N\ge 2$. For $N=2$, using $a_2= 2\, \ln 2$ from
Eq. (\ref{a2_final}), we see that the general result in Eq. (\ref{Fu0.1})
for $N=2$ matches perfectly with the leading asymptotic behavior as $u\to 0$
derived in Eq. (\ref{F2u_asymp}) from the exact scaling function for $N=2$.

\subsection{The limit $u\to 1$}

We now consider the $u\to 1$ limit of the scaling function
${\cal F}(u,N)$ in Eq. (\ref{Fu_exact.2}). It is easier to first consider
the denominator $J_N(u)$ defined in Eq. (\ref{Jnu.1}). Setting $u=1$, we find
\begin{equation}
J_N(1)= \frac{1}{2^{N-1}}\, \int_0^{\infty} dx\, e^{-x}\, \left(1-e^{-x}
\right)^{N-1}= \frac{1}{N\, 2^{N-1}}\, .
\label{Jn_u1.1}
\end{equation}
The last integral is trivially performed upon the change of variable 
$z=1-e^{-x}$.
In fact, setting $u=1-\epsilon$, it is easy to compute the next order term
for small $\epsilon$. One finds that
\begin{equation}
J_N(1-\epsilon)= \frac{1}{N\, 2^{N-1}} + b_N\, \epsilon + 
O\left(\epsilon^2\right)\, ,
\quad {\rm where}\quad b_N= -\frac{1}{2^{N-1}}\, 
\int_0^1 dz\, z^{N-1}\, \ln (1-z)\, .
\label{Jn_u1.2}
\end{equation}
We will see later that the $O(\epsilon)$ term in $J_N(1-\epsilon)$ does not
contribute to the first two leading behaviors of ${\cal F}(u,N)$ as $u\to 1$. We now turn to the integral $I_N(u)$ defined in Eq (\ref{Inu.1}). 
Setting $u=1-\epsilon$ we get
\begin{equation}
I_N(1-\epsilon)= W_N(\epsilon)=\int_0^{\infty} \frac{dx}{x^2}\, \left[1- \frac{1}{2} e^{-(1-\epsilon)\, x}
-\frac{1}{2}\, e^{-\epsilon\, x}\right]^{N}\, .
\label{In_u1.1}
\end{equation}
Now, exactly at $u=1$, i.e., at $\epsilon=0$, the integral is convergent
and we get
\begin{equation}
I_N(1)= W_N(0)= \frac{a_N}{2^N}\, , \quad {\rm where} \quad
a_N= \int_0^{\infty} \frac{dx}{x^2}\left(1- e^{-x}\right)^{N}\, .
\label{In_u1.2}
\end{equation}
Note that the same $a_N$ also appeared in Eq. (\ref{In0.1}) near
the limit $u\to 0$ which is not surprising given the fact that
the function $I_N(u)$ defined in Eq. (\ref{Inu.1}) is symmetric around $u=1/2$.

Our goal is to evaluate the next order correction to $I_N(1-\epsilon)
=W_N(\epsilon)$
as $\epsilon\to 0$. To do this, we first differentiate $W_N(\epsilon)$
in Eq. (\ref{In_u1.2}) once with respect to $\epsilon$. This gives
\begin{equation}
W_N'(\epsilon)= \frac{N}{2}\int_0^{\infty} \frac{dx}{x}\, \left[1-
\frac{1}{2}\, e^{-(1-\epsilon)\, x}- \frac{1}{2}\, e^{-\epsilon\, x}
\right]^{N-1}\, \left[ e^{-\epsilon\, x}- e^{-(1-\epsilon)\, x}\right]\, .
\label{WD1.1}
\end{equation}
We note that as $\epsilon\to 0$, the first derivative $W_N'(\epsilon)$
also diverges. To extract this divergence, we take one more derivative
of $W_N'(\epsilon)$ in Eq. (\ref{WD1.1}) with respect to $\epsilon$.
This gives
\begin{eqnarray}
W_N''(\epsilon)&= & \frac{N(N-1)}{4}\, \int_0^{\infty} dx\, \left[1-
\frac{1}{2}\, e^{-(1-\epsilon)\, x}- \frac{1}{2}\, e^{-\epsilon\, x}   
\right]^{N-2}\, \left[ e^{-\epsilon\, x}- e^{-(1-\epsilon)\, x}\right]^2 \\
&- &\frac{N}{2}\, \int_0^{\infty} dx\, \left[1-
\frac{1}{2}\, e^{-(1-\epsilon)\, x}- \frac{1}{2}\, e^{-\epsilon\, x}
\right]^{N-1}\, \left[ e^{-\epsilon\, x}+ e^{-(1-\epsilon)\, x}\right]\, .
\label{WD2.1}
\end{eqnarray}
Next we make a change of variable $\epsilon\, x=y$. This gives
\begin{eqnarray}
W_N''(\epsilon)&= & \frac{N(N-1)}{4\, \epsilon}\, \int_0^{\infty} dy\, 
\left[1-
\frac{1}{2}\, e^{- y}- \frac{1}{2}\, e^{-\frac{(1-\epsilon)}{\epsilon}\, y}
\right]^{N-2}\, \left[ e^{-y}- e^{-\frac{(1-\epsilon)}{\epsilon}\, y}
\right]^2 \\
&- &\frac{N}{2\, \epsilon}\, \int_0^{\infty} dy\, \left[1-
\frac{1}{2}\, e^{-y}- \frac{1}{2}\, e^{-\frac{(1-\epsilon)}{\epsilon}\, y}
\right]^{N-1}\, \left[ e^{-y}+ e^{-\frac{(1-\epsilon)}{\epsilon}\, y}\right]\, .
\label{WD2.2}
\end{eqnarray}
At this stage, we can now take the $\epsilon\to 0$ limit inside the integral
and neglect the term $e^{-\frac{1-\epsilon}{\epsilon}\, y}$ everywhere,
since the resulting integrals remain convergent.
This leads to the leading $\epsilon\to 0$ behavior of the second derivative
\begin{equation}
W_N''(\epsilon)\approx 
\frac{N(N-1)}{4\, \epsilon}\, \int_0^{\infty} dy\, e^{-2\, y}\, 
\left[1-
\frac{1}{2}\, e^{- y}
\right]^{N-2}
- \frac{N}{2\, \epsilon}\, \int_0^{\infty} dy\, e^{-y}\, \left[1-
\frac{1}{2}\, e^{-y}
\right]^{N-1}\, .
\label{WD2.3}
\end{equation}
These two integrals can now be easily performed by making the change of variable
$1-\frac{1}{2}\, e^{-y}=z$. This gives a very simple expression for
the leading behavior as $\epsilon\to 0$
\begin{equation}
W_N''(\epsilon)\approx - \frac{N}{2^N\, \epsilon} \, .
\label{WD2_final}
\end{equation}

Integrating back once with respect to $\epsilon$ gives
\begin{equation}
W_N'(\epsilon)\approx -\frac{N}{2^N}\, \ln(\epsilon) + C_3\, ,
\label{WD1_final}
\end{equation}
where $C_3$ is an integration constant whose value is
not relevant for us.
Integrating further with respect to $\epsilon$ we get the two leading order
terms as $\epsilon\to 0$
\begin{equation}
I_N(1-\epsilon)=W_N(\epsilon)= 
W_N(0)-\frac{N}{2^N}\, \epsilon\, \ln (\epsilon) + O(\epsilon)\, .
\label{IN_u1.3}
\end{equation}
Using $W_N(0)$ from Eq. (\ref{In_u1.2}), we then obtain, as $u\to 1$, 
\begin{equation}
I_N(u) = \frac{a_N}{2^N}\, - \frac{N}{2^N}\, (1-u)\, \ln(1-u) + O(1-u)\, ,
\label{In_u1.4}
\end{equation}
where $a_N$ is given in Eq. (\ref{In_u1.2}).

Finally, substituting the leading behaviors of $I_N(u)$ from Eq. (\ref{In_u1.4})
and of $J_N(u)$ from Eq. (\ref{Jn_u1.2}) as $u\to 1$ into our main formula
(\ref{Fu_exact.2}), we obtain the following asymptotic behavior
of ${\cal F}(u,N)$ as $u\to 1$ for arbitrary $N\ge 2$
\begin{equation}
{\cal F}(u,N)= a_N- N\, (1-u)\, \ln(1-u) + O(1-u)\, .
\label{Fu1_asymp.1}
\end{equation}

Hence, summarizing, the two limiting asymptotic behaviors of the
scaling function ${\cal F}(u,N)$, for arbitrary $N\ge 2$, are given by
\begin{eqnarray}
\label{FuN_asymp-11}
{\cal F}(u,N) \approx \begin{cases}
&\frac{a_N}{2^N-1}\, \frac{1}{u} \quad\quad\quad\quad\quad\quad\quad\quad\quad
\quad\quad\quad\quad\quad\quad {\rm as}\quad u\to 0\, \\
\\
& a_N- N\, (1-u)\, \ln(1-u) + O(1-u)\,\quad
\quad{\rm as}\quad u\to 1 \, ,
\end{cases}
\end{eqnarray} 
where we recall that the constant $a_N$ is defined in
Eq. (\ref{In_u1.2}). Note that for $N=2$, using
$a_2= 2\, \ln 2$ from Eq. (\ref{a2_final}), the
general result in Eq. (\ref{FuN_asymp-11}) does coincide
with the leading asymptotic
behaviors of ${\cal F}(u,2)$ in Eq. (\ref{F2u_asymp}) obtained
from the exact scaling function for $N=2$.

\section{Derivations of $ \langle {\mathcal{N}}_{\text{TR}} (u,N) \rangle, \langle \overline{\mathbb{T}}_N^{\text{TR}}(u) \rangle$ and $ \langle \overline{\mathbb{T}}_N^{\text{FP}}(u) \rangle$}
The MFPT under TR mechanism as in Eq. (4) of the main text can be interpreted as a sum over stochastic timescales emanating from all possible events eventually leading to a completion of the process. Following Eq. (11) of the main text, we recall
\begin{align}
 \langle  \overline{\mathcal{T}}_N^\text{TR}(u) \rangle = \langle {\mathcal{N}}_{\text{TR}} (u,N) \rangle \times \langle \overline{\mathbb{T}}_N^{\text{TR}}(u) \rangle+\langle \overline{\mathbb{T}}_N^{\text{FP}}(u) \rangle, 
\end{align}
where individual quantities were defined in the main text. In this section, we explicitly derive these quantities and elaborate further. 

\subsection{ Derivation of $ \langle {\mathcal{N}}_{\text{TR}} (u,N) \rangle$} 
\begin{figure}
    \centering
    \includegraphics[width=8cm]{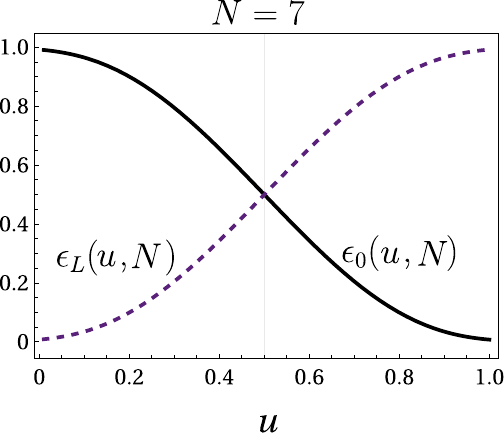}
    \caption{Variation of the splitting probabilities with $u$ for $N=7$ as found from \eref{spl-o}. Note that when $u<1/2$ we have $\epsilon_0>\epsilon_L$.}
    \label{eo}
\end{figure}

A stochastic trajectory under TR mechanism can experience multiple resetting events before the first passage to the target. The number of these TR events fluctuates between the realizations and our aim is to compute the mean number of resetting events till the first passage. 
Since, the system can undergo only two possible events: TR events (which can be called as a \textit{unsuccessful} event) and a first passage event (a \textit{successful} event), clearly the number of TR events is distributed according to a geometric distribution. Here, the \textit{success probability} is nothing but the splitting probability to the threshold  $\epsilon_L (u,N)$ so that the \textit{unsuccess probability} is given by $\epsilon_0 (u,N)$ which is the splitting probability to the target.  Thus,
the mean number of TR events can be obtained as the mean of the geometric distribution
\begin{align}
    \langle {\mathcal{N}}_{\text{TR}} (u,N) \rangle &= \sum_{k=0}^\infty k (\epsilon_L (u,N))^k \times \epsilon_0 (u,N)) =\frac{ 1-\epsilon_0 (u,N)}{ \epsilon_0 (u,N)}, \label{reset-no}
\end{align}
where  $\epsilon_0 (u,N)$ can be found using \eref{exit-o} as follows
\begin{align}
\epsilon_0 (u,N)=1-\epsilon_L (u,N)=   N u\int_0^{\infty} d\tau~ \frac{e^{-\frac{u}{\tau }} }{2 \tau ^2} \left(1-\frac{1}{2} e^{-\frac{1-u}{\tau }}-\frac{e^{-\frac{u}{\tau }}}{2}\right)^{N-1}. \label{spl-o}
\end{align}
In Fig. (\ref{eo}) we show the variation of the splitting probabilities with $u$ for $N=7$. Note that for $u<1/2$ we have $\epsilon_0>1/2$. In Fig. (\ref{ntr})(a), we plot $\langle {\mathcal{N}}_{\text{TR}} (u,N) \rangle$ as a function of $u$ for several values of $N$.

\subsection{ Derivation of $ \langle \overline{\mathbb{T}}_N^{\text{TR}}(u) \rangle$} 
\begin{figure}[t!]
    \centering
    \includegraphics[width=16.4cm]{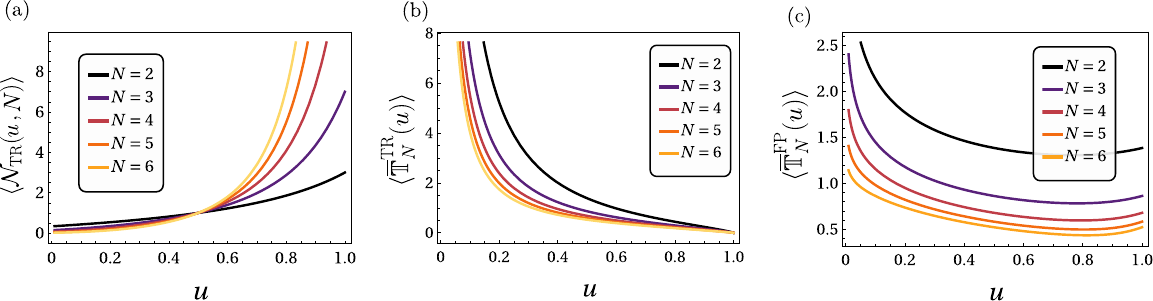}
    \caption{Variation of $ \langle {\mathcal{N}}_{\text{TR}} (u,N) $ as in \eref{reset-no}, $   \langle \overline{\mathbb{T}}_N^{\text{TR}}(u)\rangle$ as in \eref{mean-tr} and $ \langle \overline{\mathbb{T}}_N^{\text{FP}}(u) \rangle$ as in \eref{mean-tf} with respect to $u$ for distinct values of $N$.  }
    \label{ntr}
\end{figure}

Recall that the random variable ${\mathbb{T}}_N^{\text{TR}}$ denotes the waiting time between two consecutive TR events conditioned on survival. The normalized density of this conditional time can then be written as
\begin{align}
    f_{{{\mathbb{T}}_N^{\text{TR}}}}(t)&=\frac{j_{L,N}(L,x_0,t)}{\epsilon_L(L,x_0,N)}\nonumber\\
   & =\frac{ j_{L,1}(L,x_0,t) [Q(L,x_0,t)]^{N-1} }{\int_0^\infty dt~ j_{L,1}(L,x_0,t) [Q(L,x_0,t)]^{N-1}},\label{tlden}
\end{align}
where the common factor $N$ has been canceled out from the numerator and denominator. From the distribution, one can find the mean time between two consecutive resetting events as 
\begin{align}
    \langle {\mathbb{T}}_N^{\text{TR}} (L,x_0)\rangle=\int_0^\infty dt~t  f_{{{\mathbb{T}}_N^{\text{TR}}}}(t). \label{meanr}
\end{align}
Plugging the values of the survival probability as in \eref{qu} and the probability current to the threshold as in \eref{jlu} in the above equation one arrives at $\langle {{\mathbb{T}}_N^{\text{TR}}} \rangle$ (in a dimensionless form)
  \begin{align}
    \langle \overline{\mathbb{T}}_N^{\text{TR}}(u)\rangle&= \frac{v_0}{x_0} \langle{{\mathbb{T}}_N^{\text{TR}}} (L,x_0) \rangle =\frac{ \int_0^\infty dt~ \frac{ 1}{2 t}\exp \left({-\frac{1-u }{t}}\right)\left[1-\frac{1}{2} \exp \left({-\frac{1-u }{t}}\right)-\frac{1}{2}\exp\left({-\frac{ u}{t}}\right)\right]^{N-1}}{u\int_0^\infty dt~\frac{ 1}{2 t^2}\exp \left({-\frac{1-u}{t}}\right)\left[1-\frac{1}{2} \exp \left({-\frac{1-u }{t}}\right)-\frac{1}{2}\exp\left({-\frac{ u}{t}}\right)\right]^{N-1}}.  \label{mean-tr}
\end{align}
In Fig. Fig. (\ref{ntr})(b), we plot $ \langle \overline{\mathbb{T}}_N^{\text{TR}}(u)\rangle$ as a function of $u$ for different $N$.

\subsection{Derivation of $ \langle \overline{\mathbb{T}}_N^{\text{FP}}(u) \rangle$} Finally, we turn our attention to the random variable ${\mathbb{T}}_N^{\text{FP}}$ that denotes the random time for the first passage since the last TR event. This is simply the conditional time that the target is found before the threshold by any of the $N$ searchers. Following these arguments, we can write the normalized density of ${\mathbb{T}}_N^{\text{FP}}$ as
\begin{align}
    f_{{{\mathbb{T}}_N^{\text{FP}}}}(t)&=\frac{j_{0,N}(L,x_0,t)}{\epsilon_0(L,x_0,N)}\nonumber\\
   & =\frac{ j_{0,1}(L,x_0,t) [Q(L,x_0,t)]^{N-1} }{\int_0^\infty dt~ j_{0,1}(L,x_0,t) [Q(L,x_0,t)]^{N-1}}\label{toden}.
\end{align}
Finally plugging the survival probability \eref{qu} and the probability current to the target as in \eref{jou} we find the mean of  ${\mathbb{T}}_N^{\text{FP}}$ (in the dimensionless form) as  
\begin{align}
    \langle \overline{\mathbb{T}}_N^{\text{FP}}(u) \rangle&= \frac{v_0}{x_0} \int_0^{\infty} dt~tf_{{\mathbb{T}}_N^{\text{FP}}}(t) =\frac{ \int_0^\infty dt~ \frac{ 1}{2 t}\exp \left({-\frac{u }{t}}\right)\left[1-\frac{1}{2} \exp \left({-\frac{1-u }{t}}\right)-\frac{1}{2}\exp\left({-\frac{ u}{t}}\right)\right]^{N-1}}{u\int_0^\infty dt~\frac{ 1}{2 t^2}\exp \left({-\frac{u}{t}}\right)\left[1-\frac{1}{2} \exp \left({-\frac{1-u }{t}}\right)-\frac{1}{2}\exp\left({-\frac{ u}{t}}\right)\right]^{N-1}}.  \label{mean-tf}
\end{align}
In Fig. (\ref{ntr})(c), we plot $ \langle \overline{\mathbb{T}}_N^{\text{FP}}(u)\rangle$ as a function of $u$ for different values of $N$.

\subsection{Asymptotic behavior in the limit $u \to 1$}
In the following, we present the asymptotic forms of the above-mentioned observables that was used in the main text to explain the behavior of MFPT in the limit of $u \to 1$. For instance, the splitting probability to the target, in this limit, takes the following form
\begin{align}
    \epsilon_0(u=1,N)= N \int_0^{\infty} d\tau~ \frac{e^{-\frac{1}{\tau }} }{2 \tau ^2} \left(\frac{1}{2} -\frac{e^{-\frac{1}{\tau }}}{2}\right)^{N-1}=\frac{1}{2^N}.
\end{align}
Thus, from \eref{reset-no}, we have
\begin{align}
    \langle {\mathcal{N}}_{\text{TR}} (u=1,N) \rangle=2^N -1.
\end{align}
The mean conditional time between TR events, however, in this limit vanishes to zero as can be verified by setting $u=1$ in \eref{mean-tr} as seen below
      \begin{align}
    \langle \overline{\mathbb{T}}_N^{\text{TR}}(u\to 1) \rangle& =\frac{ \int_0^\infty dt~ \frac{ 1}{ t}\left[1-\exp\left({-\frac{ 1}{t}}\right)\right]^{N-1}}{\int_0^\infty dt~\frac{ 1}{ t^2}\left[1-\exp\left({-\frac{ 1}{t}}\right)\right]^{N-1}} \to 0.
    \label{Mean-u-1}
\end{align}
On the other hand, 
the conditional mean first passage time to the target $ \langle \overline{\mathbb{T}}_N^{\text{FP}}(u) \rangle$ has a finite value as can be seen from \eref{mean-tf} by setting the limit $u \to 1$ as follows 
\begin{align}
    \langle \overline{\mathbb{T}}_N^{\text{FP}}(u\to 1) \rangle= N\int_0^\infty dt~ \frac{ 1}{ t}\exp \left({-\frac{1 }{t}}\right)\left[1 -\exp\left({-\frac{ 1}{t}}\right)\right]^{N-1},
\end{align}
which was used to analyze the form of the MFPT in the $u \to 1$ limit.

\section{Diffusive searchers under TR}
\label{eg-diffusion}
In this section, we extend the discussion on the first-passage properties of diffusive search under TR. The formalism prescribed in the main text can be directly used to compute the statistical properties for this process as well. We start by recalling the single searcher diffusive propagator $G(x,t|x_0)$ in the presence of two absorbing boundaries at $x=0$ and $x=L$ from literature \cite{redner2001}
\begin{align}
    G(x,t|x_0)=\frac{2}{L} \sum _{n=0}^{\infty}  \sin \left(\frac{\pi  n x}{L}\right) \sin \left(\frac{\pi  n x_0}{L}\right) e^{-\frac{n^2 \pi ^2 D t}{L^2}}, \label{single-prop}
\end{align}
where $D$ is the diffusion constant. Next, we would need to compute the single searcher survival probability $Q(L,x_0,t)$, the flux through the origin $ j_{0,1}(L,x_0,t)$ and the threshold $j_{L,1}(L,x_0,t)$ respectively which can be obtained using the following relations
\begin{align}
   & Q(L,x_0,t)=\int_0^L G(x,t|x_0)dx, \\
    & j_{0,1}(L,x_0,t)=D\left.\frac{\partial G(x,t|x_0)}{\partial x}\right|_{x=0}, \\
     &j_{L,1}(L,x_0,t)  =-D\left.\frac{\partial G(x,t|x_0)}{\partial x}\right|_{x=L}.
\end{align}
\noindent
Using the exact expression of the propagator for the diffusive searchers, we find
\begin{align}
    &Q(u,t)=\frac{2}{\pi}\sum _{n=1}^{\infty} \left(\frac{1-(-1)^n }{  n} \right)\sin \left(n\pi u\right) e^{-\frac{n^2 \pi ^2 t}{\tau_d}},
    \label{surv-nd}\\
    & j_{0,1}(u,t)=\frac{2 \pi }{\tau_d} \sum _{n=1}^{\infty} n \sin  \left(n\pi u\right) e^{-\frac{n^2 \pi ^2  t}{\tau_d}},\\
  & j_{L,1}(u,t) =-\frac{2 \pi}{\tau_d} \sum _{n=1}^{\infty} n (-1)^n \sin  \left(n\pi u\right) e^{-\frac{n^2 \pi ^2 t}{\tau_d}}, \label{jlnd}
\end{align}
where
\begin{align}
    u=\frac{x_0}{L}, ~~~~~ \tau_d=\frac{L^2}{D}.
\end{align}
Evidently, the dimensionless MFPT under TR should become a function of the parameters $u$ and $N$, so that 
\begin{align}
 \langle   \overline{\mathcal{T}}_N^{\text{TR}}(u) \rangle= \frac{D}{x_0^2}\langle \mathcal{T}_N^{\text{TR}}(L,x_0) \rangle=\mathcal{F}(u,N),   
\end{align}
with $\mathcal{F}(u,N)$ being the scaling function. Substituting the relevant observables for the single searcher as obtained in Eq. (\ref{surv-nd})-(\ref{jlnd}) into Eq. (4) of the main text, we obtain the exact expression of the scaling function $\mathcal{F}(u,N)$ given by
\begin{align}
\mathcal{F}(u,N)
&=\frac{1}{u^2}\left(\frac{ \int_0^\infty dt~ \left[\frac{2}{\pi}\sum _{n=1}^{\infty} \left(\frac{1-(-1)^n }{  n} \right)\sin \left(n\pi u\right) e^{-n^2 \pi ^2 t}\right]^N}{N\int_0^\infty dt~ \left[2 \pi  \sum _{n=1}^{\infty} n \sin  \left(n\pi u\right) e^{-n^2 \pi ^2  t}\right] \left[\frac{2}{\pi}\sum _{n=1}^{\infty} \left(\frac{1-(-1)^n }{  n} \right)\sin \left(n\pi u\right) e^{-n^2 \pi ^2 t}\right]^{N-1}}\right). \label{mfpt-diff-11}
\end{align}

\begin{figure}[H]
    \centering
     \includegraphics[width=17cm]{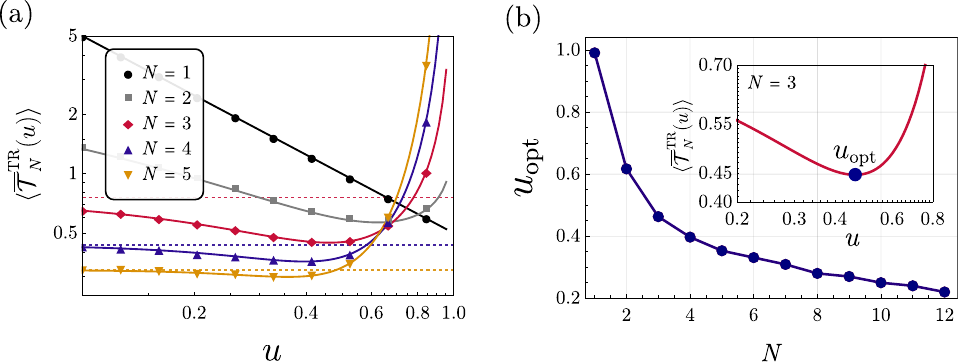}
     \caption{(a) Variation of the non-dimensionalized MFPT $\langle   \overline{\mathcal{T}}_N^{\text{TR}}(u) \rangle$ as a function of $u=x_0/L$. The solid lines represent the analytical result as in \eref{mfpt-diff-11} and the markers represent the simulation results for different values of number of searchers $N$. The dashed horizontal lines represent the reset-free case \textit{i.e.} when $L\to \infty$ or $u\to 0$. Note that for $N=1$ and $N=2$ the respective dashed lines are not shown since the MFPT of the underlying process in these cases are infinity. In the simulation, we have fixed $x_0=0.5$ and varied $L$. The value of the diffusion constant was set at $D=1$. (b) Shows the variation of the point $u_{\text{opt}}$ (inset shows an example for the case of $N=3$ searchers) with respect to $N$. Note that, there are two possible ways to realize both the figures \fref{fig2}(a,b) physically. Either one can fix $x_0$ and vary $L$ from $[x_0,\infty]$ or one can fix $L$ and vary $x_0$ from $(0,L)$. Both the cases, although, physically different, will indeed yield the same curves as shown in the above figures. This is because, the dimensionless MFPT in \eref{mfpt-diff-11} does not depend explicitly on $x_0$ or $L$, but only on the ratio $u=x_0/L$. From \fref{fig2}(b) we see that $u_{\text{opt}}\to 0$ as $N \to \infty$.}
     \label{fig2}
 \end{figure}

Through the above change of variables, we have recast the dimensionless mean first-passage time (MFPT), $\langle \overline{\mathcal{T}}_N^{\text{TR}}(u) \rangle$, as a function of a single dimensionless parameter $u \in (0,1)$, effectively eliminating dependence on the original parameters $L$, $x_0$, and $D$. While the expression in Eq. \eqref{mfpt-diff-11} does not admit a closed-form solution for arbitrary $N$, it can be evaluated numerically with high precision using Mathematica, yielding results in excellent agreement with numerical simulations (\fref{fig2}a). In what follows, we examine in detail the behavior of the scaled MFPT—i.e., the scaling function $\mathcal{F}(u, N)$—as a function of both $u$ and $N$, and highlight the physical insights that emerge from this analysis.

\subsection{Optimization with respect to threshold} \label{var-threshold}

As before, we ask the following question: Can the variation of $u$ lead to a minimal MFPT? To explore this, we plot the MFPT as a function of $u$ in Fig. \ref{fig2}(a) for various values of $N$. Remarkably, for all $N \geq 1$, the MFPT exhibits a clear minimum at an optimal value of $u$. Let us discuss these cases in the following.\\

\noindent
\textbf{\textit{Single diffusive searcher $(N=1)$}:} For a single diffusive searcher, from \eref{mfpt-diff-11} the MFPT is found to be 
\begin{align}
   & \langle \mathcal{T}_1^{\text{TR}}(L,x_0) \rangle = \frac{\frac{L^2}{2D}(x_0/L)(1-x_0/L)}{1-x_0/L}=\frac{x_0 L}{2D} \nonumber\\
   \text{or, ~} &  \langle \overline{\mathcal{T}}_1^{\text{TR}}(u) \rangle =\frac{D}{x_0^2}\langle \mathcal{T}_1^{\text{TR}}(L,x_0) \rangle=\frac{1}{2u}. \label{mfpt-n1}
\end{align}
Here, the numerator of $ \langle \mathcal{T}_1^{\text{TR}}(L,x_0) \rangle$ is the unconditional MFPT for a single Brownian searcher to reach either the boundary at $x=0$ or 
$x=L$ \cite{redner2001}. The denominator is the splitting probability for it to reach the target \cite{redner2001}. From \eref{mfpt-n1} it is evident that the MFPT for a single diffusive searcher decreases as $u^{-1}$ as also shown in \fref{fig2}(a). It is the lowest when $L\to x_0$ (i.e., $u\to 1$) where $ \langle \overline{\mathcal{T}}_1^{\text{TR}}(u=1) \rangle =1/2$. Clearly here $u \to 1$ is the optimal point as also observed in \fref{fig2}(b). Physically, such behavior can be explained in the following way:  The threshold effectively biases the searcher's motion towards the target by resetting it from $x_0$ whenever it goes away from the target. Keeping $x_0$ fixed, as one decreases $L$, therefore increasing $u$, the chances of the searchers wandering away from the target also diminish, resulting in a lower MFPT.\\\\
\textbf{\textit{Multiple diffusive searchers $(N\ge 2)$}:}
The MFPT exhibits quite distinct features for $N\ge 2$ than that for $N=1$. In here, the MFPT curves show a non-monotonic behaviour with respect to $u$ as seen in \fref{fig2}(a). 

The limit $u\to 0$ is just the reset-free case where the threshold is kept at infinity (assuming $x_0$ to be fixed). In this case, the MFPT $\langle \overline{\mathcal{T}}_N^{\text{TR}}(u\to 0) \rangle$ is just the mean fastest first passage time out of $N$ searchers to reach the target in the absence of the threshold. After some simplifications, the MFPT in this limit can be found to be
\begin{align}
    \langle \overline{\mathcal{T}}_N^{\text{TR}}(u\to 0) \rangle=\frac{1}{2} \int_0^\infty dy ~y~\left[\text{erf}\left(\frac{1}{y}\right)\right]^N, \label{lim-u-0}
\end{align}
which is a finite quantity only for $N\ge 3$. The value of the MFPT in the limit ($u\to 0$) as in \eref{lim-u-0} is shown by dashed horizontal lines in \fref{fig2}(a) for $N\ge 3$ (recall that for $N \le 2$, the underlying search has infinite MFPT and thus TR naturally expedites). As one increases $u$, the MFPT starts to decrease, showing a minimum at some intermediate value $u=u_{\text{opt}}$ so that $\left.\frac{\partial  \langle \overline{\mathcal{T}}_N^{\text{TR}}(u) \rangle}{\partial u}\right|_{u=u_{\text{opt}}}=0$ (explicitly shown in the inset of \fref{fig2}(b) for $N=3$) before increasing again as $u$ approaches unity.  As stressed earlier,  $u_{\text{opt}}$  can be thought of as the analog of optimal resetting rate in the traditional resetting set-up. In \fref{fig2}(b) we show the variation of $u_{\text{opt}}$ with respect to $N$. For $N=1$ we find $u_{\text{opt}}=1$ since the lowest MFPT occurs only at $u=1$ as evident from \eref{mfpt-n1}. However, for any other values of $N\ge 2$ we find $0<u_{\text{opt}}<1$.  

For intermediate values of $u$, the MFPT curves fall below their respective dashed lines—each representing the reset-free case for different $N$—clearly indicating that threshold resetting enhances search efficiency [\fref{fig2}(a)]. Although it is known that the MFPT for diffusive searchers is finite only for $N \geq 3$ \cite{lindenberg1980lattice,krapivsky2010maximum}, the TR mechanism enables a further reduction by optimally constraining the searchers' trajectories. This efficiency gain arises from the presence of a finite threshold that prevents searchers from diffusing beyond the latter. However, as $u \to 1$, the MFPT begins to increase again for all $N \geq 2$, in contrast to the monotonic behavior observed for $N = 1$. In this regime, the initial position $x_0$ is very close to the threshold $L$, resulting in a high splitting probability $\epsilon_L$ of reaching the threshold rather than the target. Consequently, frequent resets disrupt the search process. As $N$ increases, the likelihood that at least one searcher reaches the threshold before the target also increases, leading to a suppression of successful target-hitting events and a corresponding rise in the MFPT.

\section{First-passage time density under TR}
In this section, we present a detailed study of the FPT density of the collective search in the absence and presence of threshold resetting. The large time behavior of the FPT density becomes particularly useful to understand and compare the first-passage time moments between the processes.\\

\noindent
\subsection{FPT distribution of the ballistic searchers without TR ($u\to 0$)} 
\begin{figure}[H]
    \centering
    \includegraphics[width=16cm]{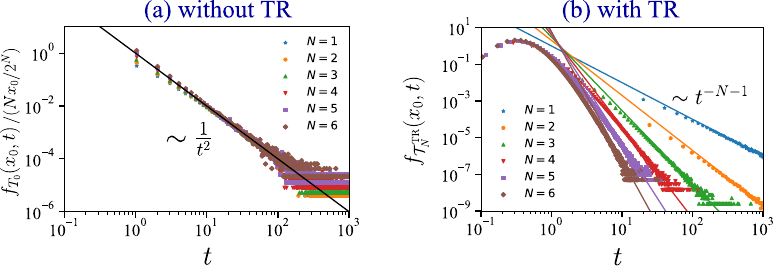}
    \caption{(a) The scaled FPT density  $ f_{T_0}(x_0,t)/\frac{N x_0}{2^{N}}$ for the ballistic searchers (without TR) exhibits $\sim 1/t^2$ decay for any values of $N$. The markers represents results from numerical simulation. The solid black line is the asymptotic result as in Eq. (\ref{asym}). (b) The FPT density $f_{\mathcal{T}_N^{\text{TR}}}(L=x_0,t)$ under optimal TR. Note that the tail exponent becomes $N$-dependent in this case as mentioned in the main text -- see Eq. (\ref{asymp-fpt-tr-II}).}
    \label{fig-den}
\end{figure}

The survival probability for a single ballistic searcher (starting from $x_0$ and target at $x=0$) in absence of the threshold is given by
\begin{align}
    Q_0(x_0,t)&= \int_0^\infty dv~ \phi(v) +  \int_{-\infty}^0 dv ~\phi(v)~ \theta(x_0 - |v| t) \nonumber\\
    &=\frac{1}{2} + \int_0^{\frac{x_0}{t}} dv ~\phi(v)\nonumber\\
    &=\frac{1}{2} + \Phi\left(\frac{x_0}{t}\right).
\end{align}
Thus, the survival probability for $N$ independent ballistic searchers are given by
\begin{align}
    Q_0^N(x_0,t)=\left[ Q_0(x_0,t)\right]^N=\left[\frac{1}{2} + \Phi\left(\frac{x_0}{t}\right)\right]^N,
\end{align}
which can be used to deduct the FPT distribution as follows
\begin{align}
    f_{T_0}(x_0,t)&=-\frac{\partial Q_0^N(x_0,t)}{\partial t}\nonumber\\
    &=N\left[\frac{1}{2} + \Phi\left(\frac{x_0}{t}\right)\right]^{N-1} \left[\frac{x_0}{t^2}\phi\left(\frac{x_0}{t}\right)\right].
\end{align}
For the exponential velocity distribution i.e., $\phi(v)=\frac{1}{2 v_0}e^{-v/v_0}$ we have $\Phi(v)=\frac{1}{2}(1-e^{-v/v_0})$. Finally plugging this expressions into the above results (setting $v_0=1$ without loss of generality) we have
\begin{align}
    f_{T_0}(x_0,t)=N\left(1-\frac{1}{2}e^{-x_0/t}\right)^{N-1}\left(\frac{x_0}{2t^2}e^{-x_0/t}\right).
\end{align}
In the limit $t\to \infty$ it has the following asymptotic behavior
\begin{align}
    f_{T_0}(x_0,t\to \infty)\sim \frac{N x_0}{2^{N}t^2}. \label{asym}
\end{align}
The exponent of $t$ at asymptotic limit is 2 (i.e., $N$ independent) leading to a diverging MFPT for any given values of $N$. The variation of the FPT density with $t$ is shown in Fig. \ref{fig-den}(a) which shows an excellent agreement with numerical simulations. \\

\noindent
\subsection{FPT distribution of the ballistic searchers under optimal TR} We have noted from the main text that the optimal MFPT occurs for the ballistic case when $u=1$ so that $L=x_0$. In here, we provide an exact derivation for the FPT distribution of the ballistic searchers at this optimality. To this end, let us recall the renewal equation for the survival probability as in Eq. (1) of the main text
\begin{eqnarray}
\mathcal{Q}_N^\text{TR}(L,x_0,t)= 
   q_N(L,x_0,t)+\int_{0}^t dt' j_{L,N}(L,x_0,t') \mathcal{Q}_N^\text{TR}(L,x_0,t-t').~~~~
\end{eqnarray}
At the optimal case of $u =1$ or $L=x_0$, the particles can survive in two possible ways. First is the the 
no-resetting case in which the survival probability is simply $q_N(L=x_0)=[Q(L,t)]^N$ (also see the main text). In the other possible scenario, one or multiple resetting events can happen. This implies that at least one of the searchers will attain a positive velocity so that resetting will surely take place -- the probability of such occurrence is given by $1-\frac{1}{2^N}$. Since the resetting here takes no time to occur as $u=1$, the probability current is simply given $j_{L,N}(L=x_0,t')=\left(1-\frac{1}{2^N}\right)\delta(t')$, where $\delta(t')$ is the Dirac-delta function. Combining both these possibilities, the renewal equation for the survival probability at $L=x_0$ takes the form
\begin{align}
    \mathcal{Q}_N^\text{TR}(L=x_0,t)=[Q(L,t)]^N + \left(1-\frac{1}{2^N}\right)\mathcal{Q}_N^\text{TR}(L=x_0,t).
\end{align}
Rearranging, we obtain an exact expression for the survival probability at the optimality i.e. $L=x_0$ as
\begin{align}
     \mathcal{Q}_N^\text{TR}(L=x_0,t)=[2 Q(L,t)]^N,
\end{align}
which is a very interesting result since it holds
for arbitrary symmetric velocity distribution of the ballistic searchers. From Eq. (5) of the main text we find that the single particle survival probability $Q(L=x_0,t)$ for a symmetric velocity distribution is given by
\begin{align}
    Q(L,t)=\Phi\left(\frac{L}{t}\right),
\end{align}
where $\Phi(v)=\int_0^v dv \phi(v)$.
Thus the survival probability has the final form 
\begin{align}
     \mathcal{Q}_N^\text{TR}(L=x_0,t)=\left[2\Phi\left(\frac{L}{t}\right)\right]^N.
\end{align}
Recall that, for the exponential velocity distribution i.e., $\phi(v)=\frac{1}{2 v_0}e^{-v/v_0}$ we have $\Phi(v)=\frac{1}{2}(1-e^{-v/v_0})$. Thus the survival probability in this case has the form (setting $v_0=1$ without any loss of generality)
\begin{align}
     \mathcal{Q}_N^\text{TR}(L=x_0,t)=\left(1-e^{-L/t}\right)^N, \label{surv-TR}
\end{align}
from which
the FPT distribution can be obtained via the relation $f_{\mathcal{T}_N^{\text{TR}}}(L=x_0,t)=-\frac{\partial \mathcal{Q}_N^\text{TR}(L=x_0,t)}{\partial t}$ and is given by
\begin{align}
    f_{\mathcal{T}_N^{\text{TR}}}(L=x_0,t)=\frac{NL}{t^2}e^{-L/t}\left(1-e^{-L/t}\right)^{N-1} , \label{asymp-fpt-tr}
\end{align}
with the following asymptotic behavior
\begin{align}
    f_{\mathcal{T}_N^{\text{TR}}}(L=x_0,t) \sim N L^N t^{-N-1} ~~ \text{as}~~ t\to \infty. \label{asymp-fpt-tr-II}
\end{align}
Thus the distribution of the FPT distribution under optimal TR for the ballistic walkers with exponential velocities follows a power tail $\sim t^{-N-1}$. This exponent makes sure that the MFPT is finite immediately when $N\ge 2$. This result is verified against numerical simulations and is shown in Fig. \ref{fig-den}(b). \\

\noindent
\textbf{Alternative derivation: } One can also derive the FPT distribution as in Eq. (\ref{asymp-fpt-tr}) but starting from Eq. (2) of the main text. This derivation, however, relies upon the exponential velocity distribution of the ballistic searchers. Let us recall Eq. (2) in the following
\begin{align}
     \widetilde{\mathcal{Q}}_N^\text{TR}(L,x_0,s)=\frac{\int_0^\infty dt~e^{-s t} [Q(L,x_0,t)]^N}{1-\int_0^\infty dt~e^{-s t}N j_{L,1}(L,x_0,t) [Q(L,x_0,t)]^{N-1}}. \label{surv-I}
\end{align}
For the ballistic walkers with exponential velocity distribution we earlier derived 
\begin{align}
    &Q(L,x_0,t)=1-\frac{1}{2}\left[e^{-x_0/t}+e^{-(L-x_0)/t}\right] \\
    &j_{L,1}(L,x_0,t)=\frac{L-x_0}{2t^2}e^{-(L-x_0)/t}.
\end{align}
In terms of these quantities the denominator can be written as
\begin{align}
    \mathcal{D}\equiv 1-N \int_0^\infty dt~e^{-s t}N\frac{L-x_0}{2t^2}e^{-(L-x_0)/t} \left[1-\frac{1}{2}(e^{-x_0/t}+e^{-(L-x_0)/t}) \right]^{N-1}.
\end{align}
To extract the limit $L\to x_0$ we first rescale $t$ as $t =(L-x_0)y$. In terms of the rescaled variable $y$ the integral can be written as 
\begin{align}
    \mathcal{D}=1-\frac{N}{2} \int_0^\infty \frac{dy}{y^2}e^{-1/y} e^{-s(L-x_0)y}\left[1-\frac{1}{2}e^{-\frac{x_0}{(L-x_0)y}}-\frac{1}{2}e^{-1/y}\right]^{N-1}.
\end{align}
Now we can take the limit $L\to x_0$ to have
\begin{align}
    \mathcal{D}=1-\frac{N}{2} \int_0^\infty \frac{dy}{y^2}e^{-1/y} e^{-s(L-x_0)y}\left[1-\frac{1}{2}e^{-1/y} \right]^{N-1},
\end{align}
that can be exactly solved to give the result
\begin{align}
     \mathcal{D}=\frac{1}{2^N},
\end{align}
which is also the splitting probability $\epsilon_0(u=1,N)$ derived earlier [see Eq. (S70) in the SM]. Substituting $\mathcal{D}$ into Eq. (\ref{surv-I}), we have
\begin{align}
    \widetilde{\mathcal{Q}}_N^\text{TR}(L=x_0,s)=2^N\int_0^\infty dt~e^{-s t} [Q(L=x_0,t)]^N=\int_0^\infty dt~e^{-s t} [1-e^{-L/t}]^N,
\end{align}
which is already in the form of a Laplace transform. Thus, by inspection, the survival probability in time domain can be written as
\begin{align}
      \mathcal{Q}_N^\text{TR}(L=x_0,t)=\left(1-e^{-L/t}\right)^N,
\end{align}
which is Eq. (\ref{surv-TR}) as derived earlier.\\


\noindent
\subsection{FPT distribution of the diffusive searchers in the absence  ($u\to 0$) and presence of TR} 
\begin{figure}[H]
    \centering
    \includegraphics[width=8cm]{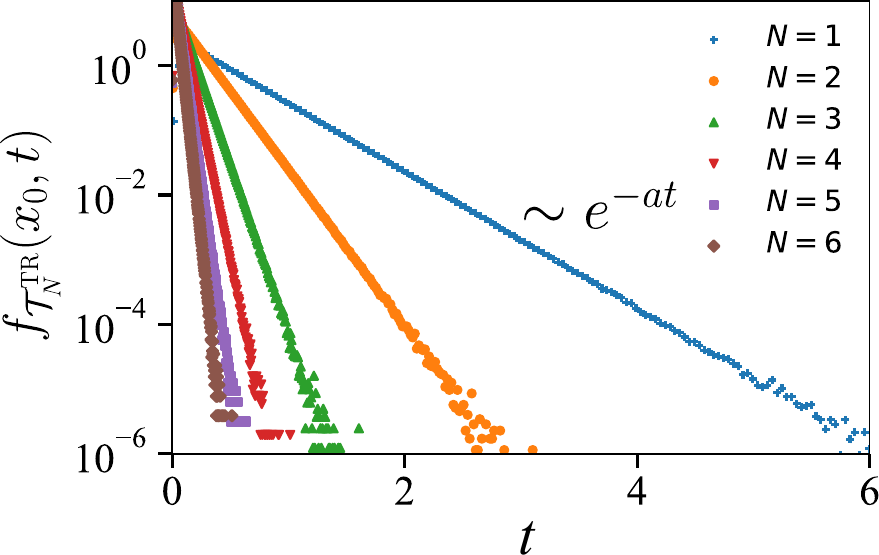}
    \caption{Numerical FPT density for diffusion with TR for various $N$ demonstrating an exponential decay.}
    \label{fig:enter-label}
\end{figure}

For diffusive searchers, the FPT  distribution is known to have the following asymptotic behavior in the absence of TR \cite{lindenberg1980lattice,krapivsky2010maximum} 
\begin{align}
    f_{T_0}(x_0,t\to \infty)\sim t^{-1-\frac{N}{2}}.
\end{align}
Note that the exponent is $N$ dependent in this case and renders an finite MFPT when $N> 2$. However, as the TR is introduced, the FPT densities for various $N$ at their respective optimal $u_\text{opt}$ are shown to exhibit exponential tails $\sim e^{-a t}$, ensuring finite moments--refer to \fref{fig:enter-label}. \\


\section{Variation of cost function with respect to $\beta$}
\begin{figure}[t!]
    \centering
    \includegraphics[width=16cm]{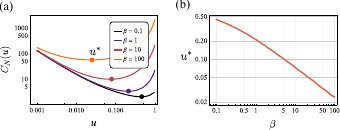}
    \caption{Panel (a): Variation of the cost function with respect to $u$ for different $\beta$ for the ballistic search case. The curves maintain non-monotonicity with a finite optimal $u^*$. Panel (b): Dependence of optimal $u^*$ as a function of $\beta$.}
    \label{cost-2}
\end{figure}

\begin{figure}[t]
    \centering
    \includegraphics[width=8cm]{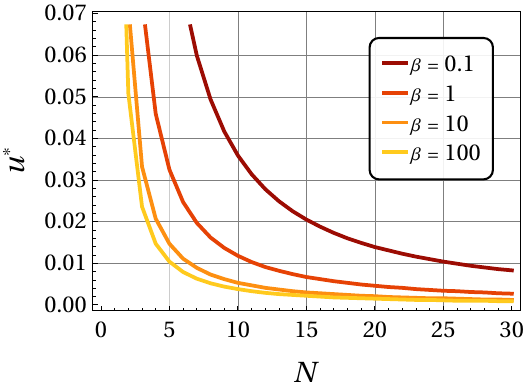}
    \caption{Variation of $u^*$ (where the cost function is optimal) as a function of $N$ for different values of $\beta$.}
    \label{figuvn}
\end{figure}

We recall the definition of cost-function from the main text
\begin{align}
    C_N (u)=\langle \overline{\mathcal{T}}_N^\text{TR}(u) \rangle + \beta N  \langle {\mathcal{N}}_{\text{TR}} (u,N) \rangle,  
\end{align}
where
$\beta$ is a constant that can be thought of cost per resetting event for a single searcher and thus, we multiply by $N$ to enumerate the total cost for the ``reset operation''. To understand the variation of the cost function with respect to $\beta$, we plot $C_N(u)$ for different $u$ and find that the curves exhibit non-monotonic dependence on $u$ for any $\beta$ -- see \fref{cost-2}(a). It is further noted that $u^*$ monotonically decreases as $\beta$ is varied to larger values -- see \fref{cost-2}(b). Overall, this study further strengthens our claim that threshold resetting can indeed optimize the collective cost regardless of the individual increase in cost. In \fref{figuvn} we show the variation of $u^*$ with $N$, but for several $\beta$. We particularly observe that, for a given $N$, optimal $u^*$ decreases as $\beta$ is increased. Physically, for a higher value of $\beta$ in this regime, the cost per resetting event becomes large so that resetting the system too frequently (i.e., increasing $u$, see \fref{ntr}(a)) will significantly increase the cost function. Thus, the optimal way for the system to function is to have a lower value of $u$ to avoid high cost due to threshold resetting.

\begin{figure}[H]
    \centering
    \includegraphics[width=8cm]{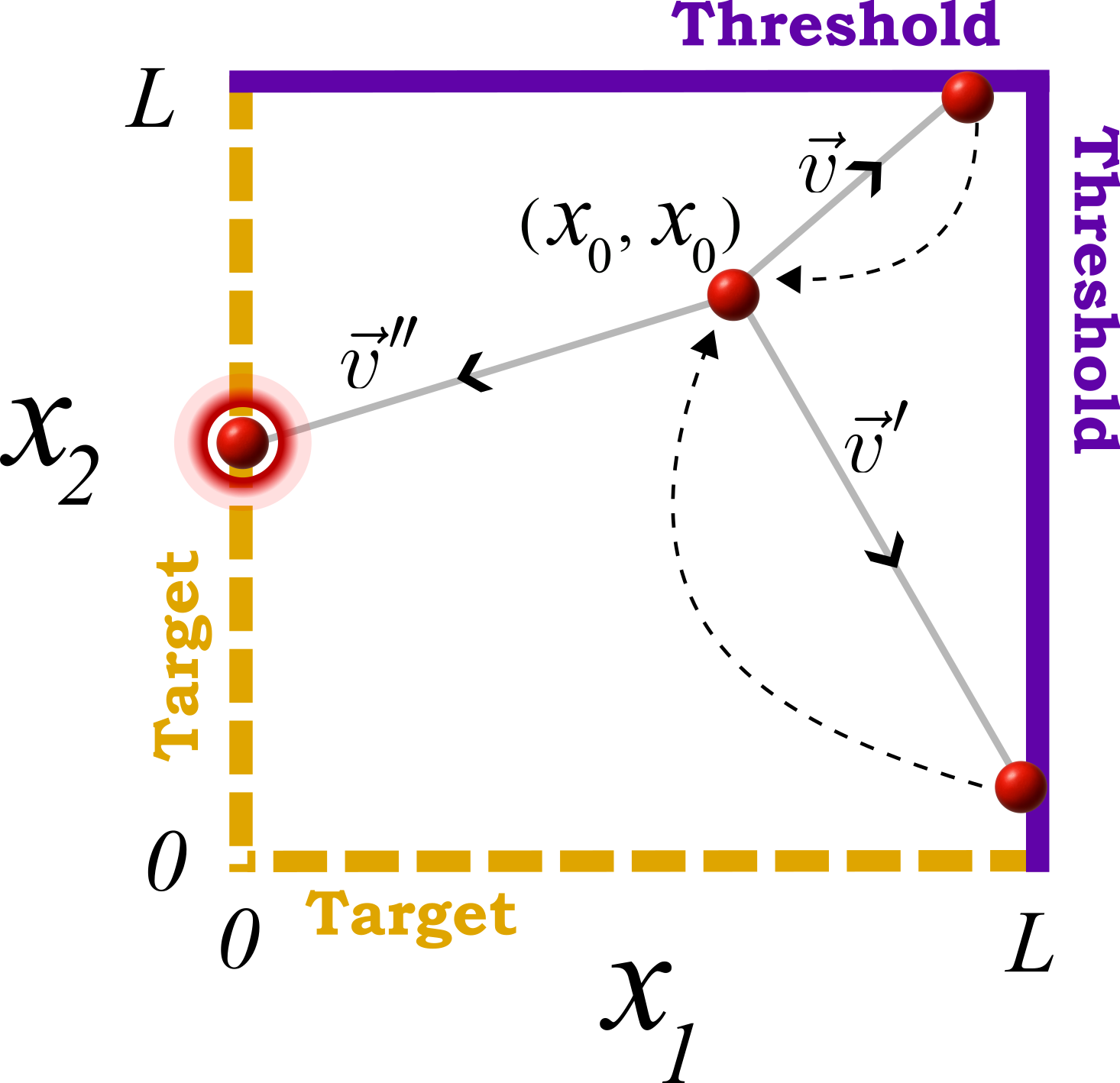}
    \caption{A higher dimension embedding depiction to realize \textit{simultaneous resetting} for two searchers in 1D by mapping to a single searcher in 2D. A single 2D searcher starts from $(x_0,x_0)$ inside a square with the target shown by the dashed lines and the resetting threshold by the solid lines. The searcher moves with certain velocity $\Vec{v}$, but as soon as it hits one of the thresholds, it is reset back to $(x_0,x_0)$. In this way, both the coordinates are reset simultaneously and new $\Vec{v'}$ is drawn. This is equivalent to 2 independent searchers getting reset at once when one of them hits the threshold and subsequently, they restart with new velocities. Here, we have shown two subsequent resetting events followed by a successful target search with a renewed velocity $\Vec{v''}$.}
    \label{nat-tr}
\end{figure}

\end{document}